\documentclass[aps,pre,twocolumn,footinbib,longbibliography]{revtex4-1}
\usepackage{amsmath}
\usepackage{amsfonts}
\usepackage{amssymb}
\usepackage{lmodern,dsfont}
\usepackage{graphicx}
\usepackage[usenames,dvipsnames]{xcolor}
\usepackage{bm}
\usepackage{blkarray}
\usepackage{blindtext}
\usepackage[english=american]{csquotes}

\newcommand{\av}[1]{\left\langle #1 \right\rangle}
\newcommand{\n}{\nonumber}
\newcommand{\nn}{\nonumber \\}
\newcommand{\mb}{\bm}
\newcommand{\dif}[1]{\frac{\text{d}}{\text{d} #1}}
\newcommand{\ind}[1]{\int \text{d} #1}
\newcommand{\Ind}[1]{\int \text{d} \bm{#1}}

\begin{document}

\author{Andreas Dechant}
\affiliation{Department of Physics \#1, Graduate School of Science, Kyoto University, Kyoto 606-8502, Japan}
\author{Shin-ichi Sasa}
\affiliation{Department of Physics \#1, Graduate School of Science, Kyoto University, Kyoto 606-8502, Japan}
\title{Entropic bounds on currents in Langevin systems}
\date{\today}

\begin{abstract}
We derive a bound on generalized currents for Langevin systems in terms of the total entropy production in the system and its environment.
For overdamped dynamics, any generalized current is bounded by the total rate of entropy production.
We show that this entropic bound on the magnitude of generalized currents imposes power-efficiency tradeoff relations for ratchets in contact with a heat bath: Maximum efficiency---Carnot efficiency for a Smoluchowski-Feynman ratchet and unity for a flashing or rocking ratchet---can only be reached at vanishing power output.
For underdamped dynamics, while there may be reversible currents that are not bounded by the entropy production rate, we show that the output power and heat absorption rate are irreversible currents and thus obey the same bound.
As a consequence, a power-efficiency tradeoff relation holds not only for underdamped ratchets but also for periodically driven heat engines.
For weak driving, the bound results in additional constraints on the Onsager matrix beyond those imposed by the Second Law.
Finally, we discuss the connection between heat and entropy in a non-thermal situation where the friction and noise intensity are state-dependent.
\end{abstract}

\maketitle

\section{Introduction}

A defining feature of an out-of-equilibrium system is a positive irreversible entropy production.
Indeed, the Second Law of thermodynamics demands that during any thermodynamic process, the total change in entropy is greater or equal than zero \cite{Fey65}.
More precisely, zero change in entropy is only possible for infinitely slow processes, during which the system is in thermal equilibrium at any point.
For most applications, however, a thermodynamic process has to occur on a finite time scale.
A useful engine, for example, should possess a finite power output.
Such a finite current---defined as the rate of change some physical observable over time--- is necessarily accompanied by a strictly positive rate of irreversible entropy production.

While the connection between entropy production and non-equilibrium is a very strong and universal statement, it is not quantitative.
The Second Law makes no prediction about the size of the irreversible entropy production or the currents in the system.
While it seems reasonable that a small rate of entropy production rate should not allow for the presence of arbitrarily large currents, this statement does not follow from the Second Law.
For a stochastic dynamics in contact with a heat bath, an explicit relation between the rate of entropy production and the size of heat currents has recently been derived by Shiraishi et al.~\cite{Shi16}: 
The square of the heat current between the system and the heat bath is bounded from above by a system-dependent positive constant times the instantaneous rate of entropy production.
This establishes that any heat current is accompanied by a minimal rate of entropy production.

A related result, originally suggested by Barato et al.~in the form of a so-called thermodynamic uncertainty relation \cite{Bar15}, has recently been proven and investigated in various contexts \cite{Gin16,Pie17,Pig17,Hor17,Dec17}.
It states that the square of the current is bounded from above by the product of the variance of the current and the entropy production.
While applicable to more general types of currents and stochastic dynamics, this relation has the drawback of being restricted to steady states and thus excludes any time-dependent driving, which is a crucial ingredient of many real-world non-equilibrium systems.
However, the similarity between the thermodynamic uncertainty relation and the bound derived in Ref.~\cite{Shi16} suggests that a similar instantaneous bound may hold for more general currents and dynamics.

In this work, we show that, for a general stochastic dynamics described by a set of Langevin equations, such a bound can be derived using the Cauchy-Schwarz inequality.
This bound generalizes the result of Ref.~\cite{Shi16} to currents other than heat currents, non-thermal heat baths and dynamics with broken time-reversal symmetry:
The square of any irreversible current is bounded from above by a positive constant times the rate of total entropy production.
We argue that this entropic bound is in fact a more precise statement of the Second Law of thermodynamics: It provides a non-zero lower bound on the rate of entropy production in terms of the square of any irreversible current.
While the derivation of the bound is mathematically straightforward, when applied to physical systems, it offers some intriguing insights into the connection between currents and entropy.

In Section \ref{sec-currents}, we specify the class of systems that will be investigated over the course of this work and define generalized currents.
In Section \ref{sec-entropy}, we then show that the irreversible part of a such a generalized current is bounded from above by the rate of entropy production.
This is the most general statement of the bound, which we will apply to specific situations in Sections \ref{sec-overdamped} through \ref{sec-nonthermal}.
As a general consequence of the bound, we establish in Section \ref{sec-entropy-connection} a connection between the system (or Shannon) entropy production and the entropy production in the medium.
For overdamped Langevin dynamics with only even variables under time-reversal, discussed in Section \ref{sec-overdamped}, any current is an irreversible current.
We argue that in this case, the entropy production rate serves as a measure of the passage of time for the macroscopic state of the system.
We apply the entropic bound on the current to derive a tradeoff relation between power and efficiency for ratchet models, generalizing the relation derived in Ref.~\cite{Shi16} for time-periodic heat engines.
This tradeoff relation states that maximal efficiency can only be realized at vanishing output power, independent of whether the ratchet is driven by a temperature difference or a time-dependent ratchet potential.
In the presence variables that are odd under a reversal of time---in particular velocities or a magnetic field---there are generally reversible currents as well as irreversible ones.
However, as we show in Section \ref{sec-under}, most currents of interest are in fact irreversible and thus bounded by the entropy production rate.
The power-efficiency tradeoff relations derived for overdamped dynamics thus also apply to the underdamped case.
One important class of model that serves as a prototype for stochastic heat engines \cite{Sch07,Bli12,Dec15,Shi16} is a trapped particle with a periodically varying temperature and trapping force.
We show that the tradeoff relation derived in Ref.~\cite{Shi16} remains valid in the presence of a magnetic field and also applies when the engine is operated in reverse, serving as a Brownian refrigerator.
Though valid arbitrarily far from equilibrium, the entropic bound also yields insights into the properties of a system close to equilibrium.
As we discuss in Section \ref{sec-onsager}, constraints on the Onsager coefficients in the linear response regime arise as a consequence of the bound.
These constraints were proven before by Brandner et al.~\cite{Bra14}, but our analysis illuminates their physical origin in a refinement of the Second Law.
Finally in Section \ref{sec-nonthermal}, we discuss systems that are in contact with a non-thermal heat bath, represented as a velocity-dependent friction and diffusion coefficient.
While in this case, the thermodynamic correspondence between heat and entropy is lost, we show that the bound remains valid and thus establishes a connection between heat and entropy in non-thermal situations. 
This leads to a tradeoff relation between power and efficiency for non-thermal engines.

\section{Generalized currents in Langevin systems} \label{sec-currents}

We consider a set of $M$ coupled Langevin equations for the dynamical variables $\bm{x}(t) = (x_1(t), \ldots, x_M(t))$
\begin{align}
\dot{x}_i(t) = A_i(\bm{x}(t),t) + \sqrt{2 B_i(\bm{x}(t),t)} \cdot \xi_i(t) \label{langevin},
\end{align}
where $i = 1, \ldots, M$, $A_i(\bm{x},t)$ and $B_i(\bm{x},t) \geq 0$ are arbitrary (time-dependent) functions of $\bm{x}$ and $\xi_i(t)$ are mutually independent Gaussian white noises, $\av{\xi_i(t) \xi_j(s)} = \delta_{i j} \delta(t-s)$.
Here, we choose to interpret the multiplicative noise as an It{\=o} product, without loss of generality, since another interpretation just renormalizes the drift coefficients $A_i(\bm{x},t)$.
We can also describe these dynamics by the Smoluchowski-Fokker-Planck equation for the probability density $P(\bm{x},t)$ and current $\bm{J}(\bm{x},t)$ \cite{Ris86}
\begin{subequations}
\begin{align}
\partial_t P(\bm{x},t) &= -\sum_{i=1}^M \partial_{x_i} J_i(\bm{x},t) \label{continuity} \\
J_i(\bm{x},t) &= \Big( A_i(\bm{x},t) - \partial_{x_i} B_i(\bm{x},t) \Big) P(\bm{x},t) \label{prob-current} .
\end{align} \label{fokkerplanck}%
\end{subequations}
The time evolution of the average $\av{Y}_t$ of some observable $Y(\bm{x},t)$ with respect to the probability density $P(x,t)$ can then be expressed in terms of the probability currents
\begin{align}
\frac{\text{d}}{\text{d}t} \av{Y}_t &= \frac{\text{d}}{\text{d}t} \int \text{d}\bm{x} \ Y(\bm{x},t) P(\bm{x},t)  \nn
&= \av{\partial_t Y}_t + \int \text{d}\bm{x} \ Y(\bm{x},t) \partial_t P(\bm{x},t) \nn
&= \av{\partial_t Y}_t + \sum_{i=1}^M \int \text{d}\bm{x} \ \big[\partial_{x_i} Y(\bm{x},t)\big] J_i(\bm{x},t)  \label{observable-evolution} ,
\end{align}
where we used the continuity equation \eqref{continuity} and integrated by parts, assuming natural boundary conditions, i.~e.~that the probability density and current vanish at the boundaries.
Here, and throughout the rest of the paper, we adopt the convention that a derivative operator inside brackets only acts in terms inside the brackets, e.~g.~$[\partial_x f] g = g \partial_x f$; by contrast, parentheses are transparent to a derivative operator, e.~g.~$(\partial_x f) g = g \partial_x f + f \partial_x g$.
Equation \eqref{observable-evolution} states that a change in the average $\av{Y}_t$ decomposes into the explicit time dependence of the function $Y(\bm{x},t)$ and the time-evolution of the dynamical variables $\bm{x}(t)$, expressed via the probability currents.
Note that this is similar to the decomposition of the energy change into work and heat employed in the framework of stochastic thermodynamics \cite{Sek10,Sei12}.
We may thus refer to the first part on the right-hand side of Eq.~\eqref{observable-evolution} as work-like and the second part as heat-like.
In this work, we will be mainly be interested in the latter, heat-like, contribution.
We define a generalized average current $\dot{R}(t)$
\begin{align}
\dot{R}(t) = \sum_{i=1}^M \int \text{d}\bm{x} \ Z_i(\bm{x},t) J_i(\bm{x},t), \label{generalized-current}
\end{align}
with arbitrary functions $Z_i(\bm{x},t)$.
The heat-like part of Eq.~\eqref{observable-evolution} is included in this definition for the particular choice $Z_i(\bm{x},t) = \partial_{x_i} Y(\bm{x},t)$, however, the above definition is more general and also includes observables that cannot be written as a total time derivative of an ensemble-averaged quantity.
Note that the definition Eq.~\eqref{generalized-current} is equivalent to the average of the generalized current discussed in Refs.~\cite{Che15,Bar15B,Dec17}.
Examples of such generalized currents are the velocities of the dynamical variables, the heat current between a particle and a heat bath or the entropy production rate (see below).

\section{Time reversal and entropic bound} \label{sec-entropy}

We assume that all variables and parameters governing the dynamics Eq.~\eqref{langevin} are either even or odd under time-reversal.
The time-reversed drift coefficients $A_i^\dagger(\bm{\epsilon} \bm{x},t)$ are obtained by reversing the sign on all odd (velocity-like) variables $\bm{\epsilon x}(t) = (\epsilon_1 x_1(t), \ldots, \epsilon_M x_M(t))$, with $\epsilon_i = 1$ for even and $\epsilon_i = -1$ for odd variables.
The sign is also reversed on all odd parameters (for example a magnetic field) that may appear in the explicit form of $A_i(\bm{x},t)$.
We further demand that the diffusion coefficients be even under time-reversal, $B_i^\dagger(\bm{\epsilon x},t) = B_i(\bm{x},t)$.
We define the reversible and irreversible probability currents via
\begin{subequations}
\begin{align}
J_i^\text{rev}(\bm{x},t) &= \underbrace{\frac{1}{2} \Big( A_i(\bm{x},t) - \epsilon_i A_i^\dagger(\bm{\epsilon x},t) \Big)}_{\equiv A_i^\text{rev}(\bm{x},t)} P(\bm{x},t) \label{rev-current} \\
J_i^\text{irr}(\bm{x},t) &= \underbrace{\frac{1}{2} \Big( A_i(\bm{x},t) + \epsilon_i A_i^\dagger(\bm{\epsilon x},t) \Big)}_{{\equiv A_i^\text{irr}(\bm{x},t)}} P(\bm{x},t) \label{irr-current}  \\
& \qquad - \partial_{x_i} B_i(\bm{x},t) P(\bm{x},t) \n .
\end{align}\label{rev-irr-current}%
\end{subequations}
We further define the entropy production during a time interval $[0,t]$ as \cite{Sei12,Spi12}
\begin{align}
\Delta S^\text{tot} &= \Delta S^\text{sys} + \Delta S^\text{med} \label{path-entropy} \\
\Delta S^\text{sys} &= \int \text{d}\bm{x} \ \Big( P(\bm{x},0) \ln P(\bm{x},0) - P(\bm{x},t) \ln P(\bm{x},t) \Big) \nn
\Delta S^\text{med} &= \int_0^t \mathcal{D}\bm{x}(s) \ \mathcal{P}(\bm{x}(s)) \ln \frac{\mathcal{P}(\bm{x}(s) \vert \bm{x}(0))}{\mathcal{P}^\dagger(\bm{\epsilon x}^\dagger(s) \vert \bm{\epsilon x}^\dagger(0))} \n .
\end{align}
The first part, the system entropy production, is the Shannon entropy difference between the initial and final state of the system.
In the second part, the medium entropy production, we integrate over all possible paths $\lbrace \bm{x}(s) \rbrace_{s \in [0,t]}$ where the path probability density $\mathcal{P}(\bm{x}(s))$ measures the probability for traversing a given path with initial conditions distributed according to $P(\bm{x},0)$. 
$\mathcal{P}(\bm{x}(s) \vert \bm{x}(0))$ is the the probability of a path given the initial condition $\bm{x}(0)$.
The time-reversed path probability density $\mathcal{P}^\dagger(\bm{x}(s) \vert \bm{x}(0))$ is obtained from the forward path probability density by (i) reversing the explicit time-dependence of all parameters and (ii) reversing the sign of all odd parameters (like a magnetic field).
Note that $\bm{x}^\dagger(s) = \bm{x}(t-s)$ denotes the time-reversed trajectory.
The path integral in Eq.~\eqref{path-entropy} can be evaluated explicitly by introducing a suitable discretization of time.
The result is the compact expression for the total entropy production \cite{Spi12}
\begin{align}
\Delta S^\text{tot}(t) = \int_0^t \text{d}s  \int \text{d}\bm{x} \ \sum_{i = 1}^M  \frac{\big(J_i^\text{irr}(\bm{x},s)\big)^2}{B_i(\bm{x},s) P(\mb{x},s)} \label{entropy-current} .
\end{align}
This shows that the total entropy production increases monotonically with time and is determined by the irreversible probability currents.
The entropy production rate $\sigma^\text{tot}(t) = \text{d}\Delta S^\text{tot}(t)/\text{d}t$ further decomposes into a sum of positive contributions
\begin{align}
\sigma^\text{tot}(t) &= \sum_{i=1}^M \sigma^\text{tot}_i(t) \label{entropy-decomp} \\
\text{with} \qquad \sigma^\text{tot}_i(t) &= \int \text{d}\bm{x} \ \frac{\big(J_i^\text{irr}(\bm{x},s)\big)^2}{B_i(\bm{x},s) P(\mb{x},s)} \n.
\end{align}
Note that a finite entropy production requires either that all the diffusion coefficients are strictly positive $B_i(\bm{x},t) > 0$ or that, if there are vanishing diffusion coefficients, the associated irreversible currents also vanish.
If the latter condition is not satisfied, this corresponds to a deterministic contraction of phase space, which leads to fully irreversible transitions and thus infinite entropy production \cite{Zer12,Rah14}.
In the following, we will assume that the entropy production is finite.

Since, by definition, the probability currents can be decomposed into the reversible and irreversible part $J_i(\bm{x},t) = J_i^\text{rev}(\bm{x},t) + J_i^\text{irr}(\bm{x},t)$, we can similarly decompose the generalized current Eq.~\eqref{generalized-current} into a reversible and an irreversible part.
\begin{align}
\dot{R}(t) &= \dot{R}^\text{rev}(t) + \dot{R}^\text{irr}(t) \label{currents-rev-irr} \\
\dot{R}^{\text{rev}/\text{irr}}(t) &= \sum_{i=1}^M \int \text{d}\bm{x} \ Z_i(\bm{x},t) J^{\text{rev}/\text{irr}}_i(\bm{x},t) . \n 
\end{align}
We now rewrite the integral over $\bm{x}$ as an average,
\begin{align}
\int \text{d}\bm{x} \ Z_i(\bm{x},t) J^{\text{irr}}_i(\bm{x},t) = \av{\frac{Z_i J_i^\text{irr}}{P}}_t,
\end{align}
and use the Cauchy-Schwarz inequality to bound
\begin{align}
\av{\frac{Z_i J_i^\text{irr}}{P}}_t^2 = \av{\frac{Z_i \sqrt{B_i} J_i^\text{irr}}{\sqrt{B_i} P}}_t^2 \leq \av{Z_i^2 B_i}_t \av{\frac{{J_i^\text{irr}}^2}{B_i P^2}}_t.
\end{align}
We identify the second factor on the right-hand side as the contribution $\sigma_i^\text{tot}(t)$ to the entropy production rate.
We thus have
\begin{align}
\bigg(\int \text{d}\bm{x} \ Z_i(\bm{x},t) J^{\text{irr}}_i(\bm{x},t)\bigg)^2 \leq \av{Z_i^2 B_i}_t \sigma_i^\text{tot}(t) .
\end{align}
Thus every single contribution to $\dot{R}^\text{irr}(t)$ is bounded by the corresponding contribution to the entropy production rate times a positive factor.
We can also use this to bound the total irreversible generalized current
\begin{align}
\Big(\dot{R}^\text{irr}(t)\Big)^2 &\leq \Bigg(\sum_{i=1}^M \bigg\vert \int \text{d}\bm{x} \ Z_i(\bm{x},t) J^{\text{irr}}_i(\bm{x},t) \bigg\vert \Bigg)^2 \nn
&\leq \Bigg(\sum_{i=1}^M \sqrt{\av{Z_i^2 B_i}_t \sigma_i^\text{tot}(t)} \Bigg)^2 ,
\end{align}
and, once more using the Cauchy-Schwarz inequality,
\begin{align}
\Big(\dot{R}^\text{irr}(t)\Big)^2 &\leq \sum_{i=1}^M \av{Z_i^2 B_i}_t \sigma^\text{tot}(t) \label{current-inequality}.
\end{align}
We have thus shown that the magnitude of an irreversible generalized current is bounded from above by the total entropy production rate in the system.
The inequality \eqref{current-inequality} constitutes the first main result of this paper.
This general bound is a more detailed statement of the Second Law of thermodynamics for stochastic dynamics:
The rate of entropy production is not only positive, but bounded from below by the square of any irreversible current in the system.

\section{Medium and system entropy} \label{sec-entropy-connection}

\begin{figure}
\includegraphics[width=.47\textwidth]{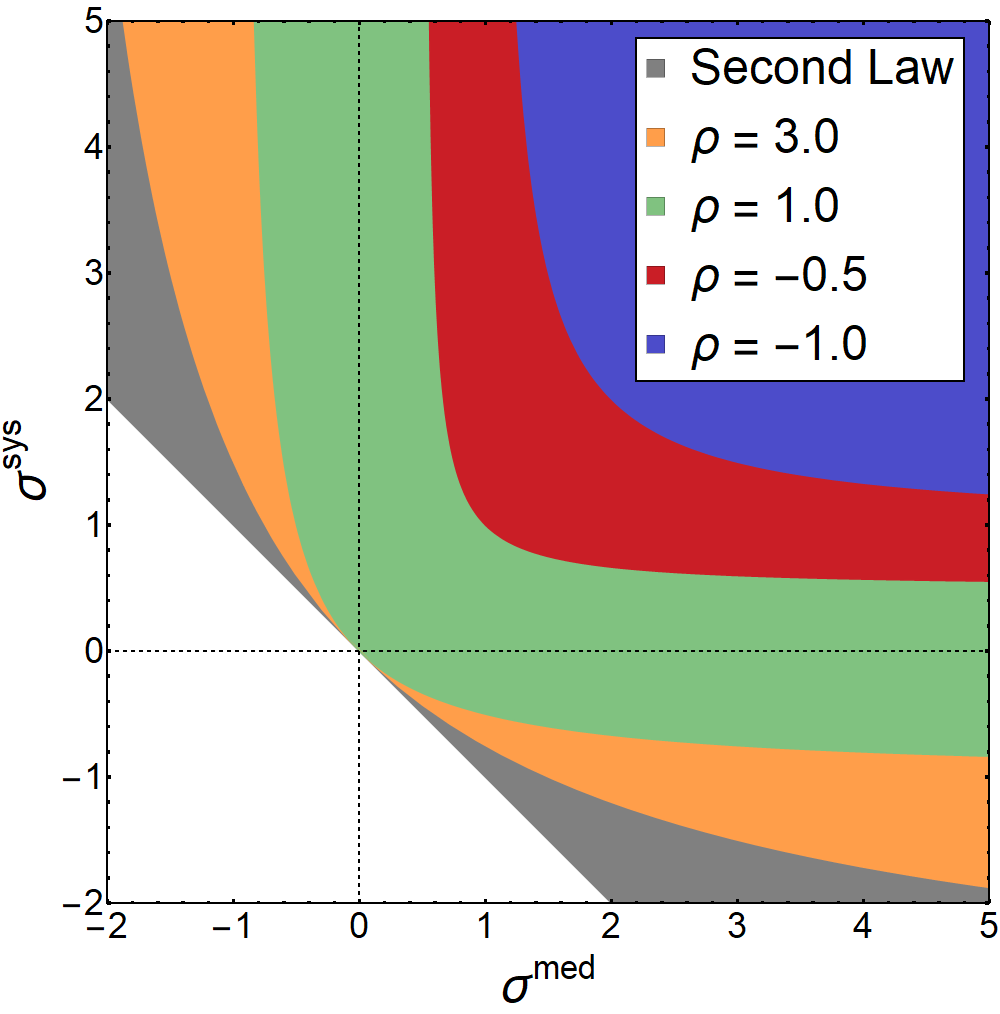}
\caption{(Color online.) Allowed values for the medium and system entropy production rates, indicated by the shaded areas. While the Second Law only requires the total entropy production rate to be positive (gray), the bound \eqref{med-sys-entropy-bound} is tighter and restricts the possible values for the medium and system contribution. For $\rho > 0$ (orange, green) an equilibrium state can exist in principle, since $\sigma^\text{med} = \sigma^\text{sys} = 0$ is permitted.
Also, for any $\sigma^\text{med} > 0$, $\sigma^\text{sys} = 0$ is allowed, and thus the possibility of a non-equilibrium steady state.
The case $\rho < 0$ (red, blue), by contrast, cannot occur in a steady state and thus corresponds to transient behavior with both medium and system entropy production rates being positive. \label{fig:entropy}}
\end{figure}

We can split the entropy production rate Eq.~\eqref{entropy-decomp} into the rates of system and medium entropy production
\begin{subequations}
\begin{align}
\sigma^\text{sys}(t) &= - \frac{\text{d}}{\text{d}t} \av{\ln P}_t \label{entropy-sys} \\
\sigma^\text{med}(t) &= \sum_{i=1}^M \av{\frac{(A^\text{irr}_i - \partial_{x_i} B_i)^2}{B_i}  + \partial_{x_i} \Big(A^\text{irr}_i - \partial_{x_i} B_i \Big)}_t, \label{entropy-med}
\end{align} \label{entropy-sys-med}%
\end{subequations}
which correspond to the time derivatives of the respective quantities in Eq.~\eqref{path-entropy}.
The system entropy production rate, which is equivalent to the rate of change in the Gibbs-Shannon entropy, vanishes in the steady state $\partial_t P(\bm{x},t) = 0$.
Since the entropy production in the medium is an irreversible current with
\begin{align}
\sigma^\text{med}(t) = \sum_{i=1}^M \int \text{d}\bm{x} \ \frac{A_i^\text{irr}(\bm{x},t)-\partial_{x_i} B_i(\bm{x},t)}{B_i(\bm{x},t)} J_i^\text{irr}(\bm{x},t),
\end{align}
we can apply the bound \eqref{current-inequality} to find
\begin{align}
\big(\sigma^\text{med}(t)\big)^2 &\leq \chi(t) \big(\sigma^\text{med}(t) + \sigma^\text{sys}(t)\big) \label{med-entropy-bound} \\
\text{with} \quad \chi(t) &= \sum_{i=1}^M \av{\frac{(A_i^\text{irr}-\partial_{x_i}B_i)^2}{B_i}}_t \geq 0 \n ,
\end{align}
where we used the decomposition $\sigma^\text{tot}=\sigma^\text{med}+\sigma^\text{sys}$.
Comparing the quantity $\chi$ with Eq.~\eqref{entropy-med}, we can write
\begin{align}
\sigma^\text{med}(t) &= \chi(t) - \rho(t) \\
\text{with} \quad \rho(t) &= -\sum_{i=1}^M \av{\partial_{x_i} A_i^\text{irr} - \partial_{x_i}^2 B_i}_t \n .
\end{align}
Plugging this into Eq.~\eqref{med-entropy-bound}, we get
\begin{align}
\rho(t) \big( \sigma^\text{med}(t) + \sigma^\text{sys}(t)\big) \geq -\sigma^\text{med}(t) \sigma^\text{sys}(t) \label{med-sys-entropy-bound}.
\end{align}
This relation implies a more intricate connection between the medium and the system entropy production than the bare Second Law $\sigma^\text{med} + \sigma^\text{sys} \geq 0$.
The relation \eqref{med-sys-entropy-bound} between the medium and system entropy production rates is shown graphically in Fig.~\ref{fig:entropy}.
In many cases, the quantity $\rho$ is positive; for example we have $\rho = \gamma$ for an underdamped particle under Stokes friction with damping rate $\gamma$, and $\rho = \kappa/(m \gamma)$ for an overdamped particle in a harmonic potential with spring constant $\kappa$ (more generally, $\rho$ is determined by average curvature of the potential).
In this case, we have
\begin{align}
\sigma^\text{med}(t) + \sigma^\text{sys}(t) \geq - \frac{\sigma^\text{med}(t) \sigma^\text{sys}(t)}{\rho(t)} .
\end{align}
While this inequality is redundant if both $\sigma^\text{med}$ and $\sigma^\text{sys}$ are positive, it becomes meaningful if the medium and system entropy production have opposite sign.
In particular, it is not possible to have $\sigma^\text{med} = -\sigma^\text{sys}$ with a finite value for $\sigma^\text{med}$, i.~e.~vanishing total entropy production requires both the system and medium part to vanish.
Moreover, the product of the two terms provides a lower bound on their sum.
Let, e.~g., be $\sigma^\text{sys} < 0$ and $\sigma^\text{med} > 0$, then
\begin{align}
\sigma^\text{med}(t) \geq \frac{\rho(t) |\sigma^\text{sys}(t)|}{\rho(t) - |\sigma^\text{sys}(t)|}
\end{align}
Thus, the rate at which the system entropy decreases is bounded by $\rho$, and the corresponding rate, at which the medium entropy increases, diverges as the former approaches this bound.
Since a decrease of system entropy corresponds to a compression of the phase space available to the system \cite{Jay65}, there is a \enquote{speed limit} for this phase-space compression and that limit can be approached only at the cost of diverging dissipation.
The case $\rho < 0$, which may occur, e.~g., for the relaxation of an overdamped particle initially located at a maximum of the potential, leads to
\begin{align}
\sigma^\text{med}(t) + \sigma^\text{sys}(t) \leq  \frac{\sigma^\text{med}(t) \sigma^\text{sys}(t)}{|\rho(t)|}
\end{align}
In this case, both the medium and system entropy production rates have to be positive; thus $\rho < 0$ cannot occur in a steady state and always corresponds to transient behavior.
Further, by minimizing the bound with respect $\sigma^\text{med}$ and $\sigma^\text{sys}$, we get the bound on the total entropy production rate
\begin{align}
\sigma^\text{tot}(t) \geq 4 |\rho(t)| .
\end{align}
For an overdamped particle in a potential, this translates to a bound on the total entropy production rate in terms of the curvature of the potential,
\begin{align}
\sigma^\text{tot}(t) \geq -\frac{4 \av{U''}_t}{m \gamma}.
\end{align}
For a particle initially located near the maximum of the potential ($\langle U'' \rangle_t < 0$), the rate of entropy production during the relaxation is thus bounded from below by the average curvature of the potential.

Finally, we note that the medium entropy production can be measured without knowledge of the explicit form of the probability density $P$, provided that the coefficients $A_i$ and $B_i$ entering the equations of motion are known.
By contrast, the system entropy production explicitly depends on the probability density and thus cannot be measured without knowing the solution of the Fokker-Planck equation \eqref{fokkerplanck}.
Nevertheless, the relation \eqref{med-entropy-bound} yields a lower bound on the system entropy production rate,
\begin{align}
\sigma^\text{sys}(t) \geq \frac{\rho(t)^2}{\chi(t)} - \rho(t) = - \frac{\rho(t)}{\chi(t)} \sigma^\text{med} \label{sys-entopy-bound-2}.
\end{align}
This lower bound on the rate of change of the Shannon entropy is expressed in terms of quantities that can be measured without knowledge about the explicit form of the probability density.
This bound is tighter than the one obtained from the Second Law, $\sigma^\text{sys} \geq -\sigma^\text{med}$.
As an example, we consider a particle under the influence of Stokes-friction and arbitrary reversible (i.~e.~velocity-independent or magnetic) external forces and in contact with a heat bath at temperature $T(t)$ (see also Section \ref{sec-under}).
In this case, the rate of entropy production in the medium is given by
\begin{align}
\sigma^\text{med}(t) = \gamma \bigg(\frac{m \av{v^2}_t}{T(t)} - 1 \bigg) = -\frac{\dot{Q}(t)}{T(t)},
\end{align}
where $\gamma$ is the damping rate, $m$ is the mass of the particle and $\langle v^2 \rangle_t$ is average of the squared velocity.
The entropy production in the medium is directly related to the heat flow $\dot{Q}$ from the heat bath to the particle.
This expression is a standard result of stochastic thermodynamics \cite{Sek10,Sei12}.
The relation Eq.~\eqref{sys-entopy-bound-2} implies that the rate of change of the Shannon entropy is bounded by the same quantities,
\begin{align}
\sigma^\text{sys}(t) \geq \gamma \bigg(\frac{T(t)}{m \av{v^2}_t} - 1 \bigg) = \frac{\dot{Q}(t)}{m \av{v^2}_t} .
\end{align}
This shows that also the system entropy production is related to the heat flow, albeit in the form of an inequality.
Since the system entropy is a total differential, we can write
\begin{align}
S^\text{GS}(t) - S^\text{GS}(0) \geq \int_0^t \text{d}s \ \frac{\dot{Q}(s)}{m \av{v^2}_s} .
\end{align}
The left-hand side is the difference in Gibbs-Shannon entropy, which depends only on the final and initial state, whereas the right-hand side measures the heat absorbed from the heat bath relative to the particle's kinetic energy, along the entire trajectory.
While the precise physical interpretation of the Gibbs-Shannon entropy for non-equilibrium states is still a matter of debate \cite{Jay65,Leb93,Gol04,Gav17}, this shows that it provides a bound on measurable currents.

\section{Overdamped dynamics} \label{sec-overdamped}

The statement of the inequality \eqref{current-inequality} is strongest for dynamics that only involve even variables and parameters under time reversal, for example for overdamped Langevin dynamics.
In this case, the reversible probability currents vanish, $J_i = J_i^\text{irr}$, and the total generalized current is bounded by entropy production rate
\begin{align}
\Big(\dot{R}(t)\Big)^2 &\leq \sum_{i=1}^M \av{Z_i^2 B_i}_t \sigma^\text{tot}(t) \label{current-inequality-even} .
\end{align}
For any explicitly time-independent observable $Y(\mb{x})$, choosing $Z_i(\bm{x}) = \partial_{x_i} Y(\bm{x})$, we find using Eq.~\eqref{observable-evolution}
\begin{align}
\bigg\vert \frac{\text{d}}{\text{d}t} \av{Y}_t \bigg\vert \leq \sqrt{\sum_{i=1}^M \av{[\partial_{x_i} Y]^2 B_i}_t \sigma^\text{tot}(t)}.
\end{align}
This means that the rate of change of any observable that does not explicitly depend on time is bounded by the rate of entropy production.
For dynamics with only even variables, any time evolution (on the level of ensemble averages) thus necessarily entails a finite entropy production rate.
Conversely, zero entropy production rate implies that there are no currents and no time evolution on average in the system.
Since the rate of entropy production bounds the time evolution of any ensemble-averaged observable, entropy can be understood as a measure of the passage of time on the ensemble level, see \cite{Leb93,Jar11} for related discussions.
Note that this intimate relation between time evolution and entropy production is lost in the presence of odd variables -- in this case zero entropy production does not preclude the presence of reversible currents and time evolution on average.

Many examples of overdamped Langevin dynamics involve periodic boundary conditions \cite{Ful75,Wea79,Mag93,Bar94,Mil94,Ast94,Rou94,Hay04,Spe06}.
Then, an additional boundary term from the integration by parts appears in Eq.~\eqref{observable-evolution},
\begin{align}
\frac{\text{d}}{\text{d}t} \av{Y}_t &= \av{\partial_t Y}_t + \sum_{i=1}^M \int \text{d}\bm{x} \ J_i(\bm{x},t) \partial_{x_i} Y(\bm{x},t) \\
& \qquad \qquad - \int_{\partial \Omega} \text{d}\bm{x} \cdot \big(Y(\bm{x},t) \bm{J}(\bm{x},t) \big), \n
\end{align}
where $\partial \Omega$ denotes the boundary of the phase-space region.
We still take Eq.~\eqref{generalized-current} as the definition of the generalized current, however, the additional boundary term means that even for $Z_i(\bm{x},t) = \partial_{x_i} Y(\bm{x},t)$ the generalized current is not equal to the time derivative of $\av{Y}_t$.
If the coordinates $x_i$ are Cartesian and we assume periodic boundary conditions for each $x_i$ individually, i.~e.~the boundary is an $M$-dimensional box, then for the choice $Z_i(\bm{x}) = \delta_{i k}$ the generalized current is the drift velocity $v_k(t)$ corresponding to $x_k$.

\subsection{Smoluchowski-Feynman ratchet} \label{sec-feynman-over}

Let us consider two paradigmatic examples: the Smoluchowski-Feynman ratchet \cite{Par96,Sek97,Mun05,Tu08} and a flashing \cite{Ajd92,Ast94} or rocking \cite{Mag93,Bar94} ratchet.
The simplest model for the Smoluchowski-Feynman ratchet, consists of two particles of masses $m_1$ and $m_2$ undergoing overdamped diffusion coupled to heat baths of temperatures $T_1$ and $T_2$.
The particles are subject to a potential $U(x_1,x_2)$, which includes interactions between the particles and external potentials, and is assumed to be spatially periodic $U(x_1+L_1,x_2) = U(x_1,x_2+L_2) = U(x_1,x_2)$.
In addition, there is a constant load force $F_0$ applied to particle $1$ against which the ratchet should perform work.
The corresponding system of Langevin equations read ($i = 1,2$)
\begin{subequations}
\begin{align}
\dot{x}_1(t) &= \frac{1}{\gamma_1} \Big( F_0 - \partial_{x_1} U(x_1(t),x_2(t)) \Big) + \sqrt{\frac{2 T_1}{\gamma_1}} \xi_1(t) \\
\dot{x}_2(t) &= -\frac{1}{\gamma_2} \partial_{x_2} U(x_1(t),x_2(t)) + \sqrt{\frac{2 T_2}{\gamma_2}} \xi_2(t),
\end{align}%
\end{subequations}
where we absorbed the masses into the damping coefficients $\gamma_i$.
The total change in position of particle $1$ over a time interval $[0,t]$ is given by
\begin{align}
x_1(t) - x_1(0) = \int_0^t \text{d}t' \ \dot{x}_1(t') 
\end{align}
or on average
\begin{align}
&\av{x_1}_t - \av{x_1}_0 \label{average-feynman} \\
& \quad = \frac{1}{\gamma_1} \int_0^t \text{d}t' \int \text{d}\bm{x} \ \Big( F_0 - \big[\partial_{x_1} U(x_1,x_2) \big] \Big) P(x_1,x_2,t') , \n
\end{align}
since the noise averages to zero.
We assume that the distribution has relaxed to a steady state $P_\text{s}(x_1,x_2)$ with the same periodicity as the potential.
Then this can be rewritten as
\begin{align}
&\av{x_1}_t - \av{x_1}_0 = t \int_0^{L_1} \text{d}x_1 \int_0^{L_2} \text{d}x_2 \ J_{\text{s},1}(x_1,x_2) ,
\end{align}
where $J_\text{s,1}$ is the $x_1$-component of the steady state probability current corresponding to $P_\text{s}$.
We thus identify the drift velocity of particle $1$,
\begin{align}
v_1 \equiv \frac{\av{x_1}_t - \av{x_1}_0}{t} = \int_0^{L_1} \text{d}x_1 \int_0^{L_2} \text{d}x_2 \ J_{\text{s},1}(x_1,x_2).
\end{align}
This has the form of a generalized current, Eq.~\eqref{generalized-current} with $Z_1 = 1$ and $Z_2 = 0$.
Then a straightforward application of Eq.~\eqref{current-inequality-even} yields the bounds
\begin{align}
\big(v_{1}\big)^2 \leq \frac{T_1}{\gamma_1} \sigma^\text{tot}_\text{s} \quad \text{and} \quad \big(v_{2}\big)^2 \leq \frac{T_2}{\gamma_2} \sigma^\text{tot}_\text{s} \label{drift-bound} .
\end{align}
The total energy of both particles is $E(t) = U(x_1(t),x_2(t)) - F_0 x_1(t)$ and we have by applying It{\=o}'s lemma
\begin{align}
\dot{E}(t) &= \big(\partial_{x_1} U(x_1,x_2) - F_0 \big) \cdot \dot{x}_1(t) + \partial_{x_2} U(x_1,x_2) \cdot \dot{x}_2(t) \nn
& \qquad + \frac{T_1}{\gamma_1} \partial_{x_1}^2 U(x_1,x_2) + \frac{T_2}{\gamma_2} \partial_{x_2}^2 U(x_1,x_2) .
\end{align}
The average change in energy is then
\begin{align}
\av{E}_t - &\av{E}_0 = \int_0^t \text{d}t' \int \text{d}\bm{x} \ \bigg[-\frac{1}{\gamma_1} \big(\partial_{x_1} U - F_0\big)^2 \\
& + \frac{T_1}{\gamma_1} \partial_{x_1}^2 U -\frac{1}{\gamma_2} \big(\partial_{x_1} U \big)^2 + \frac{T_2}{\gamma_2} \partial_{x_2}^2 U \bigg] \nn
& \qquad \times P(x_1,x_2,t') \n .
\end{align}
Again assuming a periodic steady state, we get by integrating by parts and using the fact that the potential is a periodic function of $x_1$ and $x_2$
\begin{align}
&\frac{\av{E}_t - \av{E}_0}{t}  = \dot{Q}_1 + \dot{Q}_2 \qquad \text{with} \label{heat-work-feynman} \\
\dot{Q}_1 &= \int_0^{L_1} \text{d}x_1 \int_0^{L_2} \text{d}x_2 \ \big[\partial_{x_1} U(x_1,x_2) - F_0\big] J_{\text{s},1}(x_1,x_2) \nn
\dot{Q}_2 &= \int_0^{L_1} \text{d}x_1 \int_0^{L_2} \text{d}x_2 \  \big[\partial_{x_2} U(x_1,x_2) \big] J_{\text{s},2}(x_1,x_2) \n ,
\end{align}
where we defined the heat flows $\dot{Q}_1$ and $\dot{Q}_2$ between the particles and the baths.
Integrating by parts, this can be written as
\begin{align}
\frac{\av{E}_t - \av{E}_0}{t} &= - F_0 v_1 + \int_0^{L_1} \text{d}x_1 \int_0^{L_2} \text{d}x_2 \  U(x_1,x_2)\nn
& \times \big(\partial_{x_1} J_{\text{s},1}(x_1,x_2) + \partial_{x_2} J_{\text{s},2}(x_1,x_2) \big). 
\end{align}
Since $\partial_t P_\text{s} = - \partial_{x_1}J_\text{s,1} - \partial_{x_1}J_\text{s,2} = 0$, the second term vanishes.
Then, defining the work rate $\dot{W} = -F_0 v_1$, we get the First-Law-like equality
\begin{align}
\dot{W} = \dot{Q}_1 + \dot{Q}_2 \label{first-law-feynman}.
\end{align}
A few remarks about the physical interpretation of the above definitions are in order.
Referring to the quantities $\dot{Q}_i$ as heat flows is justified in the sense of stochastic thermodynamics, in that $\dot{Q}_i$ is the ensemble average of a quantity $\dot{q}_i = -F_i \circ \dot{x}_i$, where $F_i$ is the total systematic force on particle $i$ and $\circ$ denotes a Stratonovich product.
This is the negative of the dissipation into heat bath $i$ \cite{Sek97}, i.~e.~the heat absorbed by particle $i$ from the heat bath.
The work rate $\dot{W}$ is the rate at which the particle performs work against the external load; thus, in order for the ratchet to be used as an engine, we should have $\dot{W} \geq 0$.
The heat flow $\dot{Q}_1$ has precisely the form of a generalized current with $Z_1 = \partial_{x_1} U(x_1,x_2) - F_0$ and $Z_2 = 0$, and we find the bounds
\begin{align}
\big(\dot{Q}_1 \big)^2 \leq \chi_1 T_1^2 \sigma^\text{tot}_\text{s} \quad \text{and} \quad
\big(\dot{Q}_2 \big)^2 \leq \chi_2 T_2^2 \sigma^\text{tot}_\text{s} \label{heat-bound-simple-1}  ,
\end{align}
where we defined the positive constants $\chi_1 = \av{(\partial_{x_1} U - F_0)^2}_\text{s}/(\gamma_1 T_1)$ and $\theta_2 = \av{(\partial_{x_2} U)^2}_\text{s}/(\gamma_2 T_2)$, which have dimensions of 1/time.
The inequalities \eqref{drift-bound} and \eqref{heat-bound-simple-1} state that the drift velocity of a particle as well as the heat exchange rate between the particle and the bath are bounded by the total entropy production rate.
From these inequalities, we can derive a number of useful relations.
Since the system entropy production rate is a total time derivative (see Eq.~\eqref{entropy-sys}), it vanishes in the steady state and we have $\sigma^\text{tot}_\text{s} = \sigma^\text{med}_\text{s}$ with
\begin{align}
\sigma^\text{med}_\text{s} = -\frac{1}{T_1} \dot{Q}_1 - \frac{1}{T_2} \dot{Q}_2 = \dot{Q}_1 \bigg( \frac{1}{T_2} - \frac{1}{T_1} \bigg) - \frac{\dot{W}}{T_2} \label{entropy-feynman},
\end{align}
as can be seen by evaluating Eq.~\eqref{entropy-med} explicitly and replacing $\dot{Q}_2$ using Eq.~\eqref{heat-work-feynman}.
For $\dot{W} \geq 0$, the positivity of the entropy production rate implies that either (i) $\dot{Q}_1 \geq 0$ and $T_1 \geq T_2$ or (ii) $\dot{Q}_1 \leq 0$ and $T_2 \geq T_1$.
In the following, we will focus on case (i), case (ii) follows by exchanging the labels $1$ and $2$.
The explicit expression for the heat current $\dot{Q}_1$ reads
\begin{align}
\dot{Q}_1 = - \frac{1}{\gamma_1} \av{\big(\partial_{x_1} U - F_0\big)^2}_\text{s} + \frac{T_1}{\gamma_1} \av{\partial_{x_1}^2 U}_\text{s} .
\end{align}
Since the first term on the right hand side is negative, in order to have $\dot{Q}_1 \geq 0$, we require
\begin{align}
\av{\partial_{x_1}^2 U}_\text{s} \geq \gamma_1 \chi_1 .
\end{align}
Note that this imposes a restriction on the choice of the potential function -- only a potential with specific symmetry properties will lead to a ratchet current in the desired direction.
For $T_1 \geq T_2$ and $\dot{W} \geq 0$, we define the efficiency of the ratchet as
\begin{align}
\eta = \frac{\dot{W}}{\dot{Q}_1} = \eta_\text{C} - \frac{T_2 \sigma^\text{med}_\text{s}}{\dot{Q}_1} \; \Rightarrow \; \eta_\text{C}-\eta = \frac{T_2 \sigma_\text{s}^\text{med}}{\dot{Q}_1} \label{efficiency-feyman},
\end{align}
where we introduced the Carnot efficiency $\eta_\text{C} = 1 - T_2/T_1$.
Multiplying by $\eta$ and using Eq.~\eqref{heat-bound-simple-1}, we get
\begin{align}
\dot{W} \leq \chi_1 \frac{T_1^2}{T_2}  \eta (\eta_\text{C}-\eta) \label{tradeoff-feynman}.
\end{align}
This is exactly the tradeoff relation between power and efficiency found in Ref.~\cite{Shi16}.
It states that the power produced by the ratchet vanishes as the efficiency approaches the Carnot efficiency, prohibiting Carnot efficiency at finite power.
Note that in Ref.~\cite{Shi16} the relation Eq.~\eqref{tradeoff-feynman} was derived for alternating coupling to the heat baths, whereas the Smoluchowski-Feynman ratchet is coupled to two heat baths simultaneously.
From Eqs.~\eqref{efficiency-feyman} and \eqref{heat-bound-simple-1}, we further find
\begin{align}
\sigma^\text{med}_\text{s} \leq \chi_1 \frac{T_1^2}{T_2^2} (\eta_\text{C}-\eta)^2  .
\end{align}
This provides an upper bound on the entropy production rate in terms of the efficiency of the ratchet. In particular, if the ratchet operates at Carnot efficiency, the entropy production rate is zero unless the constant $\chi_1$ diverges.
From the definition of $\chi_1$ we see that can only happen in the rather pathological case where the average square of the force on particle $1$, $\av{(\partial_{x_1} U - F_0)^2}_\text{s}$, diverges.

\subsection{Flashing and rocking ratchets} \label{sec-flashing-over}
Whereas the Smoluchowski-Feynman ratchet operates using two heat baths and a static potential, the so-called flashing \cite{Ajd92,Ast94} and rocking \cite{Mag93,Bar94} ratchets operate using a single heat bath and a potential that changes periodically in time $U(x,t+\tau) = U(x,t)$.
In the simplest case, these ratchets consist of an overdamped particle in one dimension
\begin{align}
\dot{x} = \frac{1}{\gamma} \Big(F_0 - \partial_x U(x,t) \Big) + \sqrt{\frac{2 T}{\gamma}} \xi(t) .
\end{align}
As before, the potential is also periodic in space $U(x+L,t) = U(x,t)$.
The average displacement is (see \eqref{average-feynman})
\begin{align}
&\av{x}_t - \av{x}_0 = \frac{1}{\gamma} \int_0^t \text{d}t' \int \text{d}x \ \Big[ F_0 - \partial_{x} U(x,t') \Big] P(x,t') .
\end{align}
We now assume that probability density has the same spatio-temporal periodicity as the potential $P(x,t+\tau) = P(x+L,t) = P(x,t)$.
Then, if the length of the time interval of interest is $t = n \tau$, we can write the average displacement as
\begin{align}
\av{x}_{n \tau} - \av{x}_0 = n \tau \bar{v},
\end{align}
where $\bar{v}$ is the drift velocity averaged over one period of the driving
\begin{align}
\bar{v} &= \frac{1}{\gamma \tau} \int_0^\tau \text{d}t \int_0^L \text{d}x \ \Big[ F_0 - \partial_{x} U(x,t) \Big] P(x,t) \\
&= \frac{1}{\tau} \int_0^\tau \text{d}t \int_0^L \text{d}x \ J(x,t) = L \bar{J} \n ,
\end{align}
We have from the Fokker-Planck equation
\begin{align}
\int_0^\tau \text{d}t \ \partial_{t} P(x,t) = - \partial_x \int_0^\tau \text{d}t \ J(x,t).
\end{align}
Since the left hand side is zero due to the time-periodicity of the probability density, we find that the time-averaged probability current $\bar{J} = 1/\tau \int_0^\tau \text{d}t' \ J(x,t')$ is constant in space.
From Eq.~\eqref{current-inequality-even} we find
\begin{align}
\bar{v}^2 \leq \overline{v^2} \leq \frac{1}{\tau} \int_0^\tau \text{d}t \ \frac{T}{\gamma} \sigma^\text{tot}(t) = \frac{T}{\gamma} \bar{\sigma}^\text{med} .
\end{align}
Since the system entropy production rate can be written as a total time derivative (see Eq.~\eqref{entropy-sys}), it does not contribute to the time average.
Qualitatively, we thus have the same bound on the drift velocity as Eq.~\eqref{drift-bound}, but now the steady state quantities are replaced by time averages.
For the total energy of the particle $E(t) = U(x(t),t) - F_0 x(t)$, we get
\begin{align}
\dot{E}(t) &= \partial_t U(x(t),t) + \big(\partial_x U(x(t),t) - F_0 \big) \cdot \dot{x}(t) \\
& \qquad \qquad + \frac{T}{\gamma} \partial_x^2 U(x(t),t) \n,
\end{align}
which, after averaging over the ensemble and one period of the driving, yields time-averaged rate of energy change
\begin{align}
&\frac{\av{E}_\tau - \av{E}_0}{\tau} = \dot{E}^\text{in} + \dot{Q} \quad \text{with} \label{heat-work-flashing} \\
\dot{E}_\text{in} &= \frac{1}{\tau} \int_0^\tau \text{d}t \int_0^L \text{d}x \ \big[\partial_{t} U(x,t)\big] P(x,t) \nn
\dot{Q} &= \frac{1}{\tau} \int_0^\tau \text{d}t \int_0^L \text{d}x \ \big[\partial_x U(x,t) - F_0\big] J(x,t), \n
\end{align}
where, as in the previous example, $\dot{Q}$ denotes the (now time-averaged) heat current between the particle and the heat bath.
In addition, there is now an additional energy input rate $\dot{E}^\text{in}$ due to the time-dependent driving through the variation of the potential $U(x,t)$.
Integrating by parts with respect to time, respectively space, the energy input rate and the first term in the heat current cancel and we have the connection to the time-averaged work rate $\dot{W} = -\bar{v} F_0$,
\begin{align}
\dot{W} = \dot{E}^\text{in} + \dot{Q} .
\end{align}
From Eq.~\eqref{entropy-med}, we find the time-averaged entropy production rate
\begin{align}
\bar{\sigma}^\text{med} = -\frac{\dot{Q}}{T} \label{entropy-flashing} .
\end{align}
The positivity of the entropy production now straightforwardly implies $\dot{Q} \leq 0$.
In order to have $\dot{W} \geq  0$, we thus have to demand $\dot{E}^\text{in} \geq 0$ and define the efficiency as
\begin{align}
\eta = \frac{\dot{W}}{\dot{E}^\text{in}} = 1 - \frac{T \bar{\sigma}^\text{med}}{\dot{E}^\text{in}} .
\end{align}
We then have
\begin{align}
\eta (1-\eta) = \frac{\dot{W} T \bar{\sigma}^\text{med}}{(\dot{E}^\text{in})^2} \label{eta-eq-flashing} .
\end{align}
We thus want to bound $\dot{E}^\text{in}$ by the entropy production rate.
Integrating by parts with respect to time in Eq.~\eqref{heat-work-flashing}, we get
\begin{align}
\dot{E}_\text{in} &= -\frac{1}{\tau} \int_0^\tau \text{d}t \int_0^L \text{d}x \ U(x,t) \partial_{t'} P(x,t) \nn
&=\frac{1}{\tau} \int_0^\tau \text{d}t \int_0^L \text{d}x \ U(x,t) \partial_x J(x,t) \nn
&= - \frac{1}{\tau} \int_0^\tau \text{d}t \int_0^L \text{d}x \ \big[ \partial_x U(x,t) \big] J(x,t) \label{input-current} .
\end{align}
We take the square of this expression
\begin{align}
\big(\dot{E}_\text{in}\big)^2 &= \frac{1}{\tau^2} \bigg( \int_0^\tau \text{d}t \int_0^L \text{d}x \ \big[ \partial_x U(x,t) \big] J(x,t) \bigg)^2 \nn
&\leq \frac{1}{\tau^2} \bigg( \int_0^\tau \text{d}t \bigg\vert \int_0^L \text{d}x \ \big[ \partial_x U(x,t) \big] J(x,t) \bigg \vert \bigg)^2 \nn
&\leq \frac{1}{\tau^2} \bigg( \int_0^\tau \text{d}t \sqrt{\frac{T}{\gamma}\av{\big(\partial_x U\big)^2}_{t} \sigma^\text{tot}(t')}\bigg)^2 \nn
&\leq \frac{1}{\tau} \frac{T}{\gamma} \int_0^\tau \text{d}t \ \av{\big(\partial_x U\big)^2}_{t} \ \bar{\sigma}^\text{med},
\end{align}
where we used Eq.~\eqref{current-inequality-even} from the second to the third line and the Cauchy-Schwarz inequality from the third to the fourth line.
Defining $\chi = \int_0^\tau \text{d}t' \ \av{(\partial_x U)^2}_{t'}/(\gamma \tau T)$, we thus have the bound
\begin{align}
\big(\dot{E}_\text{in}\big)^2 \leq \chi T^2 \bar{\sigma}^\text{med} .
\end{align}
Plugging this into Eq.~\eqref{eta-eq-flashing} we find a tradeoff relation similar to Eq.~\eqref{tradeoff-feynman}
\begin{align}
\dot{W} \leq \chi T \eta (1-\eta) .
\end{align}
Since the ratchet is now driven by an external variation of the potential, the efficiency is no longer bounded by the Carnot efficiency, but can reach a value of $1$.
However, as for the Smoluchowski-Feynman ratchet, reaching the maximal efficiency leads to vanishing power output.
This tradeoff relation is a consequence of two physical bounds:
On the one hand, the maximally attainable efficiency is a consequence of the Second Law of thermodynamics, which states that the entropy production rate is positive. 
This imposes a lower bound on the rate at which energy is dissipated into the heat bath, i.~e.~the rate of energy loss.
On the other hand, the vanishing power output at maximal efficiency is a consequence of Eq.~\eqref{current-inequality-even}, which bounds any current in the system, in particular also the rate of energy input, by the entropy production rate.

\section{Underdamped dynamics} \label{sec-under}

If the dynamics Eq.~\eqref{langevin} contains odd variables or parameters under time reversal, then the reversible probability currents are generally non-zero.
Then only the irreversible currents are bounded by the entropy production rate, see Eq.~\eqref{current-inequality}.
Fortunately, many currents of physical interest, in particular heat currents, turn out to be irreversible currents and thus similar statements as in the previous section are possible also for dynamics including odd degrees of freedom.
As a specific but still rather general case, we consider a set of position and velocity variables, $\bm{x} = (x_1, \ldots, x_M)$ and $\bm{v} = (v_1, \ldots, v_M)$.
These are governed by the underdamped Langevin equations
\begin{subequations}
\begin{align}
\dot{x}_i(t) &= v_i(t) \\
\dot{v}_i(t) &= \frac{1}{m_i} \Big( - \partial_{x_i} U(\bm{x}(t),t) + F_i \Big) \\
& \qquad \quad  - \gamma_i v_i(t) + \sqrt{\frac{2 \gamma_i T_i(t)}{m_i}} \xi_i(t) \n.
\end{align}
\end{subequations}
While the positions are even under time reversal, $x_i \rightarrow x_i$, the velocities are odd and change sign, $v_i \rightarrow -v_i$.
The reversible and irreversible currents are then, from the definition Eq.~\eqref{rev-irr-current},
\begin{align}
J_{x_i}^\text{rev}(\bm{x},\bm{v},t) &= v_i P(\bm{x},\bm{v},t), \qquad J_{x_i}^\text{irr}(\bm{x},\bm{v},t) = 0, \nn
J_{v_i}^\text{rev}(\bm{x},\bm{v},t) &= \frac{1}{m_i} \Big( - \big[\partial_{x_i} U(\bm{x},t)\big] + F_i \Big) P(\bm{x},\bm{v},t), \nn
J_{v_i}^\text{irr}(\bm{x},\bm{v},t) &= -\gamma_i \Big( v_i + \frac{T_i(t)}{m_i} \partial_{v_i} \Big) P(\bm{x},\bm{v},t) \label{currents-underdamped} ,
\end{align}
assuming that the non-conservative forces $F_i$ are even under time reversal.
Since the irreversible probability currents associated with the position variables vanish, we have for the total entropy production rate
\begin{align}
\sigma^\text{tot}(t) = \sum_{i=1}^M \int \text{d}\bm{x} \int \text{d}\bm{v} \ \frac{m_i \big(J_{v_i}^\text{irr}(\bm{x},\bm{v},t)\big)^2}{\gamma_i T_i(t) P(\bm{x},\bm{v},t)}.
\end{align}
For the total energy of the system $E(t) = \sum_i m_i v_i(t)^2/2 + U(\bm{x}(t),t) - \sum_i F_i x_i(t)$ we get from It{\=o}'s lemma
\begin{align}
\dot{E}(t) &= \partial_t U(x(t),t) + \sum_{i = 1}^M \bigg( m_i v_i(t) \cdot \dot{v}_i(t) + \gamma_i T_i(t) \\
& \qquad \qquad \qquad \qquad \qquad - F_i \dot{x}_i(t) \bigg) \n.
\end{align}
This yields for the change in average energy
\begin{align}
\av{E}_t - \av{E}_0 &= \int_0^t \text{d}t' \ \big(\dot{W}(t') + \dot{Q}(t') \big) \\
\text{with} \qquad \dot{W}(t) &= \av{\partial_{t} U}_{t} \nn
\dot{Q}(t) &= \sum_{i=1}^M m_i  \int \text{d}\bm{x} \int \text{d}\bm{v} \ v_i J_{v_i}^\text{irr}(\bm{x},\bm{v},t) \n ,
\end{align}
where we assumed natural boundary conditions on the velocity degrees of freedom, i.~e.~that the probability density and its derivatives vanish as $v_i \rightarrow \pm \infty$.
Any change in the average energy of the system thus decomposes into a part that involves the explicit time-dependence of the potential and a part that is proportional to the irreversible velocity probability currents.
In the spirit of stochastic thermodynamics \cite{Sek10,Sei12}, the first part can be interpreted as work done on the system and the second part as the heat absorbed from the reservoirs.
Since the heat flow has precisely the form of an irreversible current (see Eq.~\eqref{currents-rev-irr}), we can apply the inequality \eqref{current-inequality} to get a bound on the former
\begin{align}
\Big(\dot{Q}(t)\Big)^2 \leq \sum_{i=1}^M m_i \gamma_i T_i(t) \av{v_i^2}_t \ \sigma^\text{tot}(t) .
\end{align}
A similar relation holds for the individual heat currents,
\begin{align}
\Big(\dot{Q}_i(t)\Big)^2 \leq m_i \gamma_i T_i(t) \av{v_i^2}_t \ \sigma^\text{tot}(t) \label{heat-bound-underdamped} .
\end{align}
Thus any energy exchange between the system and the coupled reservoirs in the form of heat is bounded by the total entropy production rate.
Further, we get for the medium entropy rate, Eq.~\eqref{entropy-med},
\begin{align}
\sigma^\text{med}(t) = \sum_{i=1}^M \gamma_i \bigg(\frac{m_i \av{v_i^2}_t}{T_i(t)} - 1 \bigg) = - \sum_{i=1}^M \frac{\dot{Q}_i(t)}{T_i(t)} \label{entropy-med-underdamped} .
\end{align}

\subsection{Ratchets} \label{sec-ratchets-under}
We now apply this bound to the ratchet models studied in the overdamped limit in the previous section.
For the Smoluchowski-Feynman ratchet with load force $F_1 = F_0$ (and $F_2 = 0$), the potential is time independent and we have in the steady state
\begin{align}
\frac{\av{E}_t - \av{E}_0}{t} &= \dot{Q}_1 + \dot{Q}_2 \quad \text{with} \\
\dot{Q}_i &= m_i \int\text{d}\bm{x} \int\text{d}\bm{v} \ v_i J_{\text{s},v_i}^\text{irr}(\bm{x},\bm{v}) \n .
\end{align}
We can relate this to the work $\dot{W} = -F_0 \av{v_1}_\text{s}$ performed against the load force by noting
\begin{align}
0 &= -\Ind{x} \Ind{v} \ \bigg( \sum_{i=1}^2 \frac{m_i}{2} v_i^2 + U(\bm{x}) \bigg) \nn
&\qquad \qquad \qquad \cdot \sum_{i=1}^2 \bigg( \partial_{x_i} J_{\text{s},x_i} + \partial_{v_i} J_{\text{s},v_i} \bigg) \nn
&= \dot{Q}_1 + \dot{Q}_2 - \dot{W} ,
\end{align}
where we integrated by parts and used the definitions of the heat currents.
This is exactly the same as Eq.~\eqref{first-law-feynman}.
The rest of the argument then proceeds analog to the previous section and we find the tradeoff relation
\begin{align}
\dot{W} \leq \chi_1 \frac{T_1^2}{T_2}  \eta (\eta_\text{C}-\eta),
\end{align}
with the positive constant $\chi_1 = m_1 \gamma_1 \av{v_1^2}_\text{s}/T_1$.
We note that the value of the constant $\chi_1$ is different for the overdamped and underdamped description.
In particular, in the overdamped limit $m \gamma \rightarrow \infty$, the underdamped constant diverges and does not converge to the overdamped value.
The reason for this behavior is that the velocity degrees of freedom contribute a non-vanishing heat current even in the overdamped limit, which is neglected in the overdamped description \cite{Hon00,Aro17}.

For the periodically driven rocking or flashing ratchet, we again have to consider quantities that are averaged over one period $\tau$ of the driving.
We find
\begin{align}
\frac{\av{E}_\tau - \av{E}_0}{\tau} &= \dot{E}^\text{in} + \dot{Q} \quad \text{with} \\
\dot{E}^\text{in} &= \frac{1}{\tau} \int_0^\tau \text{d}t \ \av{\partial_{t} U}_{t} \nn
\dot{Q} &= \frac{1}{\tau} \int_0^\tau \text{d}t \int_0^L \text{d}x \ind{v} \ m \gamma v J_v^\text{irr}(x,v,t) \n .
\end{align}
Again, the total change in energy is found to be equal to $\dot{W} = -F_0 \overline{\av{v}}$ with $\overline{\av{v}} = \int_0^\tau \text{d}t' \av{v}_{t'} /\tau$.
We can further relate $\dot{E}^\text{in}$ to the irreversible velocity current,
\begin{align}
\dot{E}^\text{in} &= \frac{1}{\tau} \int_0^\tau \text{d}t' \int_0^L \text{d}x \ind{v} \ \big[\partial_{t'} U(x,t')\big] P(x,v,t') \nn
&= -\frac{1}{\tau} \int_0^\tau \text{d}t \int_0^L \text{d}x \ind{v} \ U(x,t) \partial_{t} P(x,v,t) \nn
&= \frac{1}{\tau} \int_0^\tau \text{d}t \int_0^L \text{d}x \ind{v} \ U(x,t) \nn
& \hspace{1.5 cm} \times \Big( \partial_x J_x(x,v,t) + \partial_v J_v(x,v,t) \Big) .
\end{align}
The integral over the derivative of the velocity current vanishes and we have, integrating by parts,
\begin{align}
&\dot{E}^\text{in} = -\frac{1}{\tau} \int_0^\tau \text{d}t \int_0^L \text{d}x \ind{v} \ v \big[\partial_x U(x,t)\big] P(x,v,t) \nn
& \; = \frac{1}{\gamma \tau} \int_0^\tau \text{d}t \int_0^L \text{d}x \ind{v} \ \big[\partial_x U(x,t)\big] J_v^\text{irr}(x,v,t),
\end{align}
in analogy to Eq.~\eqref{input-current}.
We then find the tradeoff relation
\begin{align}
\dot{W} \leq \chi T \eta (1-\eta),
\end{align}
with $\chi = \int_0^\tau \text{d}t \ \av{(\partial_x U)^2}_{t} /(m \gamma \tau T) $.
Contrary to the Smoluchowski-Feynman ratchet, the coefficient on the right hand side of tradeoff relation is the same in both the overdamped and underdamped description.
The reason is that now, only a single thermal reservoir is present and there thus is no additional heat current due to the relaxation of the velocity degree of freedom.

\subsection{Periodically driven heat engine} \label{sec-periodic-under}
While the Smoluchowski-Feynman ratchet can be considered as a type of heat engine due to the presence of two baths at different temperatures, there is a paradigmatic type of stochastic heat engine more akin to classical heat engines, which has been realized in several experimental systems \cite{Bli12,Mar16}: a single trapped particle coupled to a heat bath \cite{Sch07,Dec15}.
In this case, both the potential and the temperature are varied as functions of time.
We consider a particle of mass $m$ in three dimensions $\bm{x} = (x_1,x_2,x_3)$ under the influence of a potential force and a magnetic field,
\begin{align}
m\dot{\bm{v}} &= - \bm{\nabla} U(\bm{x},t) + q \big(\bm{v} \bm{\times} \bm{B}(\bm{x},t) \big) - q \partial_t \bm{A}(\bm{x},t) \label{langevin-magnetic} \\
& \qquad \qquad - m \gamma \bm{v} + \sqrt{2 m \gamma T(t)} \bm{\xi} \n ,
\end{align}
where $\bm{B}(\bm{x},t)$ is the magnetic field and $\bm{A}(\bm{x},t)$ is the associated vector potential, $\bm{B}(\bm{x},t) = \bm{\nabla} \bm{\times} \bm{A}(\bm{x},t)$.
Here $\bm{\nabla} = (\partial_{x_1},\partial_{x_2},\partial_{x_3})$ and $\bm{\times}$ denotes the vector product.
The potential $U(\bm{x},t)$ may be due to an electric scalar potential, but also other potential forces like gravity.
Since the magnetic field is odd under time-reversal, so is the vector potential and its time-derivative is even.
Then the reversible and irreversible probability currents read
\begin{align}
J_{x_i}^\text{rev}(\bm{x},\bm{v},t) &= v_i P(\bm{x},\bm{v},t), \qquad J_{x_i}^\text{irr}(\bm{x},\bm{v},t) = 0, \label{currents-underdamped-magnetic}  \\
J_{v_i}^\text{rev}(\bm{x},\bm{v},t) &= \frac{1}{m} \Big( - \big[\partial_{x_i} U(\bm{x},t)\big] + q \big(\bm{v} \bm{\times} \bm{B}(\bm{x},t) \big)_i  \nn
&\hspace{1.5 cm} - q \partial_t \bm{A}(\bm{x},t) \Big) P(\bm{x},\bm{v},t), \nn
J_{v_i}^\text{irr}(\bm{x},\bm{v},t) &= -\gamma \Big( v_i + \frac{T(t)}{m} \partial_{v_i} \Big) P(\bm{x},\bm{v},t) \n .
\end{align}
The total energy of the particle is $E(t) = m \bm{v}(t)^2/2 + U(\bm{x}(t),t)$.
Applying It{\=o}'s lemma and averaging yields
\begin{align}
\av{E}_t &- \av{E}_0 = \int_0^t \text{d}t' \Bigg( \av{\partial_{t'}U}_{t'} \\
& + \sum_{i = 1}^3 \Ind{x} \Ind{v} \ \bigg( v_i \Big( - \big[\partial_{x_i} U(\bm{x},t') \big] \nn
&\qquad + q \big(\bm{v} \bm{\times} \bm{B}(\bm{x},t) \big)_i - q \big[\partial_{t'} A_i(\bm{x},t) \big] - m \gamma v_i \Big) \nn
&\quad + \gamma T(t') + \big[\partial_{x_i} U(\bm{x},t') \big] v_i \bigg) P(\bm{x},\bm{v},t) \Bigg) \n .
\end{align}
The terms involving the potential force $-\bm{\nabla} U(\bm{x},t)$ cancel.
Further, the term involving the magnetic field can be written as $\bm{v} \cdot \big(\bm{v} \bm{\times} \bm{B}(\bm{x},t) \big) = 0$ since $\bm{v} \bm{\times} \bm{B}$ is orthogonal to $\bm{v}$.
The remaining terms can be written as
\begin{align}
\av{E}_t - \av{E}_0 = &\int_0^t \text{d}t' \ \big( -\dot{W}(t') + \dot{Q}(t') \big) \\
\text{with} \quad \dot{W}(t) &= -\av{\partial_t U}_t + q \av{\bm{v} \partial_t \bm{A}}_t \quad \text{and}  \nn
\dot{Q}(t) = m \Ind{x} &\Ind{v} \ \bm{v} \bm{J}_{\bm{v}}^\text{irr}(\bm{x},\bm{v},t) \n ,
\end{align}
where we defined $\bm{J}_{\bm{v}} = (J_{v_1}, J_{v_2}, J_{v_3})$.
The work has now two contributions, one due to the time-dependence of the potential, which has the same form as in the absence of a magnetic field, and an additional term stemming from the time-dependence of the vector potential and thus the magnetic field.
Due to the absence of explicitly non-conservative forces, the energy difference is equal to the difference in internal energy $\av{E}_t - \av{E}_0 = \av{H}_t - \av{H}_0$, where $H(\bm{x},\bm{v},t) = m \bm{v}^2/2 + U(\bm{x},t)$ is the Hamiltonian of the system.
The heat flow is again of the form of a generalized irreversible current and is thus bounded by the total entropy production rate according to Eq.~\eqref{current-inequality}
\begin{align}
\big(\dot{Q}(t)\big)^2 \leq m \gamma T(t) \av{\bm{v}^2}_t \ \sigma^\text{tot}(t) . \label{heat-bound-magnetic}
\end{align}
The medium entropy Eq.~\eqref{entropy-med} production now reads
\begin{align}
\sigma^\text{med}(t) = \frac{\av{\bm{v}^2}_t}{T(t)}  - 1,
\end{align}
and thus the heat flow is related to the medium entropy production rate via Eq.~\eqref{entropy-med-underdamped},
\begin{align}
\dot{Q}(t) = - T(t) \sigma^\text{med}(t).
\end{align}
Using this, we can express both $\dot{Q}(t)$ and $\langle \bm{v}^2\rangle_t$ in terms of $\sigma^\text{med}(t)$ and obtain the inequality between total, medium and system entropy production rate
\begin{align}
\sigma^\text{tot}(t) \geq - \frac{1}{\gamma} \sigma^\text{med}(t) \sigma^\text{sys}(t) \label{entropy-inequality},
\end{align}
which corresponds to Eq.~\eqref{med-sys-entropy-bound} with $\rho = \gamma$.

We take the potential, vector potential and temperature to be periodic functions of time, $U(\bm{x},t+\tau) = U(\bm{x},t)$, $\bm{A}(\bm{x},t+\tau) = \bm{A}(\bm{x},t)$ and $T(t+\tau) = T(t)$, and assume that the probability density has the same periodicity $P(\bm{x},\bm{v},t+\tau) = P(\bm{x},\bm{v},t)$.
Since the Hamiltonian $H(\bm{x},\bm{v},t) = m \bm{v}^2/2 + U(\mb{x},t)$ is time-periodic, so is its average with respect to the time-periodic probability density, and we have
\begin{align}
0 &= \av{H}_\tau - \av{H}_0 = \int_0^\tau \text{d}t \ \dif{t'} \av{H}_{t} \nn
&=\int_0^\tau \text{d}t \ \bigg( \av{\partial_{t'} U}_{t} \\
&\quad +  \Ind{x} \Ind{v} \ \Big( \frac{m}{2} \bm{v}^2 + U(\bm{x},t) \Big) \partial_{t} P(\bm{x},\bm{v},t) \bigg) \n .
\end{align}
Using the continuity equation $\partial_t P(\bm{x},\bm{v},t) = - \sum_i \big(\partial_{x_i} J_{x_i}(\bm{x},\bm{v},t) + \partial_{v_i} J_{v_i}(\bm{x},\bm{v},t) \big)$, integrating by parts with respect to $\bm{x}$ and $\bm{v}$, respectively, and using the fact that the boundary terms vanish, we then find
\begin{align}
0 = \int_0^\tau \text{d}t \ \big( -\dot{W}(t) + \dot{Q}(t) \big),
\end{align}
which means that the energy of the system is conserved over one period of the driving.
Likewise, the system entropy does not change over one period and the time-averaged entropy production rate consists only of the medium part Eq.~\eqref{entropy-med-underdamped}
\begin{align}
\bar{\sigma}^\text{med} &= - \frac{1}{\tau} \int_0^\tau \text{d}t \ \frac{\dot{Q}(t)}{T(t)} .
\end{align}
In order to define the efficiency of the engine, we follow Ref.~\cite{Bra14} and parameterize the inverse temperature in terms of minimal and maximal temperatures $T_\text{c}$ and $T_\text{h}$,
\begin{align}
\frac{1}{T(t)} = \frac{1}{T_\text{c}} - \bigg(\frac{1}{T_\text{c}} - \frac{1}{T_\text{h}} \bigg)\phi(t) \label{temp-prot} ,
\end{align}
where $\phi(t)$ is a dimensionless periodic function with $0 \leq \phi(t) \leq 1$.
Then, expressing the heat flow as $\dot{Q} = \text{d}\av{H}_t/\text{d}t + \dot{W}$, we have
\begin{align}
&\bar{\sigma}^\text{med} =  \frac{1}{\tau} \int_0^\tau \text{d}t' \ \frac{1}{T(t')} \Big( -\dot{W}(t') - \dif{t'} \av{H}_{t'} \Big) . \nn
&= -\frac{1}{T_\text{c}} \frac{1}{\tau} \int_0^\tau \text{d}t \ \dot{W}(t) + \bigg(\frac{1}{T_\text{c}} - \frac{1}{T_\text{h}} \bigg) \frac{1}{\tau} \int_0^\tau \text{d}t \ \phi(t) \dot{Q}(t) \nn
&\equiv -\frac{1}{T_\text{c}} \overline{\dot{W}} + \bigg(\frac{1}{T_\text{c}} - \frac{1}{T_\text{h}} \bigg) \overline{\dot{Q}^\text{in}} \label{entropy-magnetic}.
\end{align}
The first term is just the time-averaged work rate divided by $T_\text{c}$.
We interpret the second term as the input heat, noting that it is only nonzero if $T(t) > T_\text{c}$.
For the special case of an instantaneous temperature change, e.~g.~$\phi(t) = 0$ for $0 \leq t < \tau_{\text{c}}$ and $\phi(t) = 1$ for $\tau_\text{c} \leq t < \tau$, we indeed find
\begin{align}
\overline{\dot{Q}^\text{in}} &= \frac{1}{\tau} \int_{\tau_\text{c}}^\tau \text{d}t \ \dot{Q}(t) = \frac{\Delta Q_{\text{h}}}{\tau},
\end{align}
which is precisely the time-averaged rate of heat exchange with the hot bath.
Now we can define the efficiency as
\begin{align}
\eta = \frac{\overline{\dot{W}}}{\overline{\dot{Q}}^\text{in}} = \eta_{\text{C}} - \frac{T_\text{c} \bar{\sigma}^\text{med}}{\overline{\dot{Q}^\text{in}}} \label{efficiency-magnetic} .
\end{align}
We can then use the bound \eqref{current-inequality} to bound $\overline{\dot{Q}^\text{in}}$,
\begin{align}
\big(\overline{\dot{Q}^\text{in}}\big)^2 &\leq \chi T_\text{c}^2 \ \bar{\sigma}^\text{med} \quad \text{with} \label{input-bound-magnetic} \\
\chi = &\frac{m \gamma }{\tau T_\text{c} } \int_0^\tau \text{d}t \ \frac{\phi^2(t)}{1-\eta_\text{C} \phi(t)} \av{\mb{v}^2}_{t} \n .
\end{align}
Multiplying Eq.~\eqref{efficiency-magnetic} by $\eta_\text{C} - \eta$, we obtain the tradeoff relation
\begin{align}
\overline{\dot{W}} \leq \chi T_\text{c} \eta \big(\eta_\text{C} - \eta \big) \label{tradeoff-magnetic} .
\end{align}
For this system, such a relation was shown to be valid in the linear response regime in Ref.~\cite{Bra14}.
The above derivation shows that the relation remains valid beyond linear response.
Further, unlike Ref.~\cite{Shi16}, where the tradeoff relation was proven for systems that are even with respect to time reversal, we explicitly allow for the presence of a magnetic field which breaks time-reversal symmetry.
The relation \eqref{tradeoff-magnetic} asserts that a microscopic heat engine that can be modeled by Eq.~\eqref{langevin-magnetic} cannot realize Carnot efficiency at finite power.
Further, the power output of the engine is bounded by
\begin{align}
\overline{\dot{W}} \leq \chi \frac{T_\text{c}}{4} \eta_\text{C}^2 ,
\end{align}
which is the maximal value of the right hand side of Eq.~\eqref{tradeoff-magnetic} with respect to $\eta$.

The trapped-particle heat engine can also be operated in reverse, serving as a refrigerator.
In this case, we rewrite the time-averaged entropy production rate as
\begin{align}
\bar{\sigma}^\text{med} = -\frac{1}{T_\text{h}} \overline{\dot{W}} - \bigg(\frac{1}{T_\text{c}} - \frac{1}{T_\text{h}} \bigg) \overline{\dot{Q}^\text{abs}},
\end{align}
where we defined the time-averaged heat absorption rate from the cold reservoir as
\begin{align}
\overline{\dot{Q}^\text{abs}} &= \frac{1}{\tau} \int_0^\tau \text{d}t \ (1-\phi(t)) \dot{Q}(t)  .
\end{align}
In contrast to the heat input rate in the case of a heat engine, this quantity is nonzero only for $T(t) < T_\text{h}$ and is given by $\Delta Q_{\text{c}}/\tau$ in the case of an instantaneous temperature change.
Noting that $\overline{\dot{W}} \leq 0$, the coefficient of performance $\xi$ is defined as
\begin{align}
\xi = - \frac{\overline{\dot{Q}^\text{abs}}}{\overline{\dot{W}}} = \xi_\text{C} + \frac{T_\text{h} T_\text{c}}{T_\text{h}-T_\text{c}} \frac{\bar{\sigma}^\text{med}}{\overline{\dot{W}}} ,
\end{align}
where $\xi_\text{C} = T_\text{c}/(T_\text{h} - T_\text{c})$ is the coefficient of performance of an ideal Carnot refrigerator.
Since we have $\overline{\dot{W}} = \overline{\dot{Q}}$, we can use the bound on the heat current \eqref{heat-bound-magnetic},
\begin{align}
\big(\overline{\dot{W}}\big)^2 &\leq \chi^\text{R} T_\text{c}^2 \bar{\sigma}^\text{med} \quad \text{with} \\
\chi^\text{R} &= \frac{m \gamma}{\tau T_\text{c}}  \int_0^\tau \text{d}t \ \frac{1}{1-\eta_\text{C} \phi(t)} \av{\mb{v}^2}_{t} \n ,
\end{align}
to get a tradeoff relation for the refrigerator
\begin{align}
\overline{\dot{Q}^\text{abs}} \leq \chi^\text{R} T_\text{c} \frac{\xi \big(\xi_\text{C}-\xi \big)}{1 + \xi_\text{C}}.
\end{align}
In analogy to the tradeoff relation between output power and efficiency for a heat engine, reaching the Carnot limit for the coefficient of performance of a refrigerator necessarily leads to vanishing cooling rate.

\section{Constraints on Onsager coefficients} \label{sec-onsager}

\begin{figure}
\includegraphics[width=0.48\textwidth]{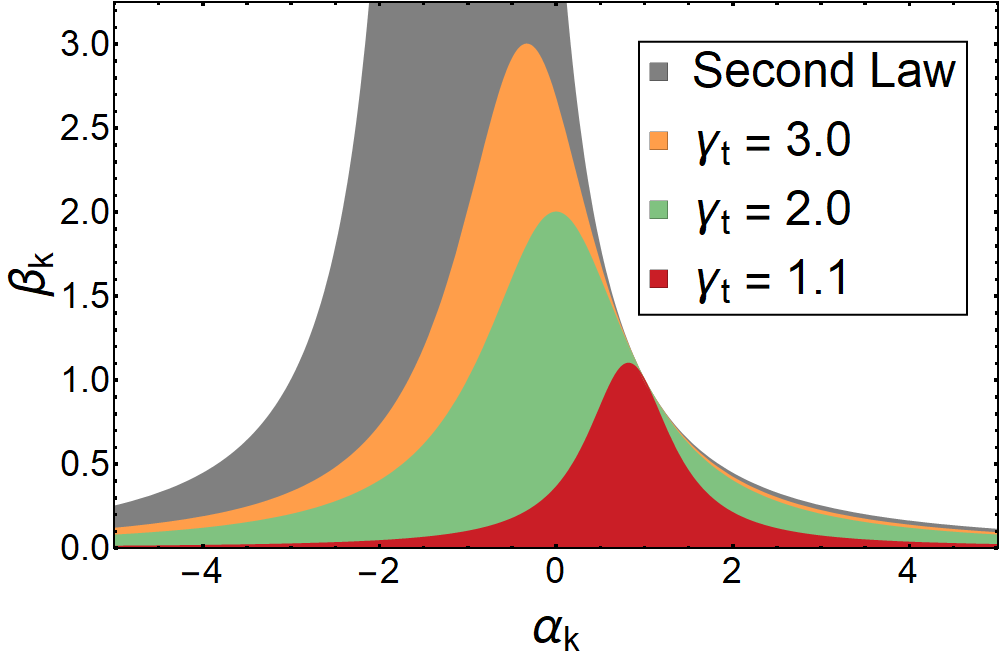} 
\caption{(Color online.) The constraint \eqref{offdiag-constraint}, for different values of $\gamma_k$ (colors), in comparison to the one obtained from the Second Law (gray). The shaded area represents the allowed values for $(\alpha_k,\beta_k)$. Note that the two constraints only coincide for $\alpha_k = 1$, i.~e.~a symmetric Onsager coefficient matrix.
In the antisymmetic case $\alpha_k = -1$, all values of $\beta_k$ are allowed by the Second Law, however, only $\beta_k \leq \gamma_t - 1$ satisfies the constraint \eqref{offdiag-constraint}.  \label{fig:onsager} }
\end{figure}

In Ref.~\cite{Bra14}, the tradeoff relation Eq.~\eqref{tradeoff-magnetic} was derived in the linear response regime by proving an additional constraint on the Onsager coefficients beyond the constraints imposed by the the Second Law of thermodynamics.
However, no interpretation of this additional constraint was given.
We are now going to show that the constraint arises naturally as a consequence of the inequality Eq.~\eqref{current-inequality}, which, as stated before, can be regarded as a more precise statement of the Second Law.
As in Ref.~\cite{Bra14}, we separate the time-dependence of the potential and magnetic field, as well as the temperature, from a time-independent equilibrium part,
\begin{subequations}
\begin{align}
U(\bm{x},t) &= U_0(\bm{x}) + \epsilon y_u U_1(\bm{x},t)\\
\bm{A}(\bm{x},t) &= \bm{A}_0(\bm{x}) + \epsilon y_a \bm{A}_1(\bm{x},t) \\ 
T_\text{h} &= T_\text{c} (1 + \epsilon y_t) ,
\end{align}
\end{subequations}
where $U_1$ and $\bm{A}_1$ are some (fixed) functions, $y_u$, $y_a$ and $y_t$ are parameters of order 1 and we take $\epsilon \geq 0$.
The linear response treatment then corresponds to assuming that $\epsilon \ll 1$ and considering only the leading order contributions to the respective quantities.
The equilibrium state $\epsilon = 0$ is most conveniently expressed in terms of the canonical momentum $\bm{p} = m \bm{v} + q \mb{A}$, and is of the Boltzmann-Gibbs form
\begin{align}
P_0(\bm{x},\bm{p}) &= Z_0^{-1} e^{-\frac{H_0(\bm{x},\bm{p})}{T_\text{c}}} \quad \text{with} \\
Z_0 &= \Ind{x} \Ind{p} \ e^{-\frac{H_0(\bm{x},\bm{p})}{T_\text{c}}} \quad \text{and} \nn
H_0(\bm{x},\bm{p}) &= \frac{1}{2 m} \big( \bm{p} - q \bm{A}_0(\bm{x}) \big)^2 + U_0(\bm{x}) \n ,
\end{align}
which is the steady state solution of the equilibrium Kramers-Fokker-Planck equation
\begin{align}
\mathcal{L}_0&(\bm{x},\bm{p}) P_0(\bm{x},\bm{p}) = 0 \quad \text{with} \\
\mathcal{L}_0(\bm{x},\bm{p}) &= - \bm{\nabla} \big[ \bm{\nabla}_{\bm{p}} H_0(\bm{x},\bm{p}) \big] + \bm{\nabla}_{\bm{p}} \big[ \bm{\nabla} H_0(\bm{x},\bm{p})\big]  \nn
& \qquad \qquad + \gamma \bm{\nabla}_{\bm{p}} \big( \bm{p} - q \bm{A}_0(\bm{x}) + m T_\text{c} \bm{\nabla}_{\bm{p}} \big) \Big) \n .
\end{align}
Assuming an expansion of the time-dependent probability density in terms of $\epsilon$
\begin{align}
P(\bm{x},\bm{p},t) = P_0(\bm{x},\bm{p}) + \epsilon P_1(\bm{x},\bm{p},t) + O(\epsilon^2),
\end{align}
we can write the time-averaged work and heat rates as
\begin{align}
\overline{\dot{W}} &= -\epsilon \Big(y_u j_u + y_a j_a \Big), \qquad \overline{\dot{Q}^\text{in}} = j_t \qquad \text{with} \label{current-relation} \\
j_u &= - \frac{\epsilon}{\tau} \int_0^\tau \text{d}t \Ind{x}\Ind{p} \ U_1(\bm{x},t) \partial_t P_1(\bm{x},\bm{p},t) \nn
j_a &= \frac{q}{m} \frac{\epsilon}{\tau} \int_0^\tau \text{d}t  \Ind{x}\Ind{p} \ \big(\bm{p}-q\bm{A}_0(\bm{x})\big) \bm{A}_1(\bm{x},t) \nn
&\hspace{3cm} \times \partial_t P_1(\bm{x},\bm{p},t) \nn
j_t &= \frac{\epsilon}{\tau} \int_0^\tau \text{d}t \ \phi(t) \Ind{x}\Ind{p} \ \bigg( H_0(\bm{x},\bm{p}) \partial_t P_1(\bm{x},\bm{p},t) \nn
&\hspace{1.5cm} + y_a \frac{q}{m} \big(\bm{p} - q \bm{A}_0(\bm{x})\big) \dot{\bm{A}}_1(\bm{x},t) P_0(\bm{x},\bm{p}) \bigg) \n ,
\end{align}
where we neglected terms of order  $\epsilon^2$ and higher in the definition of the currents $j_i$.
In terms of the currents, the time-averaged entropy production rate Eq.~\eqref{entropy-magnetic} reads
\begin{align}
\bar{\sigma}^\text{med} &= \frac{\epsilon}{T_\text{c}} \Big( y_u j_u + y_a j_a + y_t j_t \Big) + O(\epsilon^3) .
\end{align}
We can obtain a more explicit expression for the currents by using the evolution equation for the first order correction of the probability density
\begin{align}
\partial_t P_1(\bm{x},\bm{p},t) = \mathcal{L}_0(\bm{x},\bm{p}) P_1(\bm{x},\bm{p},t) + \mathcal{L}_1(\bm{x},\bm{p},t) P_0(\bm{x},\bm{p}),
\end{align}
where the first order correction to the Kramers-Fokker-Planck operator is given by
\begin{align}
\mathcal{L}_1(\bm{x},&\bm{p},t) = y_u \big[\bm{\nabla} U_1(\bm{x},t) \big] \cdot \bm{\nabla}_{\bm{p}} \\
& + y_a \frac{q}{m} \Big( \big[ \bm{\nabla} (\bm{p} \cdot \bm{A}_1(\bm{x},t)) \big] \cdot \bm{\nabla}_{\bm{p}} - \bm{A}_1(\bm{x},t)\cdot \bm{\nabla} \big] \Big) \nn
& + y_t m \gamma T_\text{c} \phi(t) \bm{\nabla}_{\bm{p}}^2 \n .
\end{align}
The formal solution to this equation reads \cite{Ris86,Bra14}
\begin{align}
P_1(\bm{x},\bm{p},t) = \int_0^\infty \text{d}s \ e^{\mathcal{L}_0(\bm{x},\bm{p}) s} \mathcal{L}_1(\bm{x},\bm{p},t-s) P_0(\bm{x},\bm{p}).
\end{align}
Since $\mathcal{L}_1$ consists of terms that are explicitly proportional to $y_u$, $y_a$ and $y_t$, we can write
\begin{align}
j_t = \epsilon T_\text{c} \big(L_{t u} y_u  + L_{t a} y_a + L_{t t} y_t \big) \label{heat-current}
\end{align}
and similar for $j_u$ and $j_a$, which defines the Onsager coefficients $L_{j k}$.
This allows us to write the entropy production rate as a quadratic form
\begin{align}
\bar{\sigma}^\text{med} = \epsilon^2 \sum_{k,l = u,a,t} y_k L_{k l} y_l = \epsilon^2 \ \bm{y} \cdot \mathbb{L} \cdot \bm{y} \label{entropy-linear-response},
\end{align}
with the matrix of Onsager coefficients $\mathbb{L}$.
Since we have $\bar{\sigma}^\text{med} \geq 0$ from the Second Law, the coefficient matrix has to be positive semi-definite.
This is equivalent to all of its principal minors having a non-negative determinant, which yields constraints on the response coefficients.
In particular, we find for any two indices $k,l$
\begin{align}
L_{k k} \geq 0, \quad L_{k k} L_{l l} - \frac{1}{4} (L_{k l} + L_{l k})^2 \geq 0 \label{onsager-bound-1} .
\end{align}
Using the bound on the heat input rate Eq.~\eqref{input-bound-magnetic} and Eq.~\eqref{current-relation} we further get the bound
\begin{align}
j_t^2 \leq \chi_t T_\text{c}^2  \bar{\sigma}^\text{med} \label{heat-bound-linear} .
\end{align}
Since in the linear response regime, we have $\eta_\text{C} \sim O(\epsilon)$, we can express the coefficient $\chi_t$ to leading order in $\epsilon$ as
\begin{align}
\chi_t &= \frac{m \gamma}{\tau T_\text{c}} \int_0^\tau \text{d}t \ \phi^2(t) \av{v^2}_0 \nn
&= \frac{\gamma}{\tau} \int_0^\tau \text{d}t \ \phi^2(t) ,
\end{align}
where we used the equilibrium average kinetic energy $m \langle v^2 \rangle_0 = T_\text{c}$.
Plugging in the expressions for $j_t$ and $\bar{\sigma}^\text{med}$, Eqs.~\eqref{heat-current} and \eqref{entropy-linear-response}, we find
\begin{align}
\big(L_{t u} y_u  + L_{t a} y_a + L_{t t} y_t \big)^2 \leq \chi_t \sum_{k, l = u,a,t} y_k L_{k l} y_l.
\end{align}
This can again be written in the form
\begin{align}
\bm{y} \cdot \mathbb{M} \cdot \bm{y} \geq 0,
\end{align}
with the new coefficient matrix
\begin{widetext}
\begin{align}
\mathbb{M} = \left( \begin{array}{ccc}
\chi_t L_{t t} - L_{t t}^2 & \frac{\chi_t}{2} \big(L_{t u} + L_{u t}\big) - L_{t u} L_{t t} & \frac{\chi_t}{2} \big(L_{t a} + L_{a t} \big) - L_{t a} L_{t t} \\[1 ex]
\frac{\chi_t}{2} \big(L_{t u} + L_{u t}\big) - L_{t u} L_{t t} & \chi_t L_{u u} - L_{t u}^2 & \frac{\chi_t}{2} \big(L_{u a} + L_{a u} \big) - L_{t u} L_{t a} \\[1 ex]
\frac{\chi_t}{2} \big(L_{t a} + L_{a t} \big) - L_{t a} L_{t t} & \frac{\chi_t}{2} \big(L_{u a} + L_{a u} \big) - L_{t u} L_{t a} & \chi_t L_{a a} - L_{t a}^2
\end{array} \right) .
\end{align}
\end{widetext}
Since the matrix has to be positive semi-definite, all of its principal minors need to have a non-negative determinant.
We find for $k = t,u,a$
\begin{align}
\chi_t L_{k k} \geq L_{t k}^2 \label{onsager-bound-2},
\end{align}
and for $k = u,a$
\begin{align}
L_{t t} \leq \chi_t \bigg(1 - \frac{1}{4} \frac{(L_{t k} - L_{k t})^2}{L_{t t} L_{k k} - L_{t k} L_{k t}} \bigg) \label{onsager-bound-3} .
\end{align}
For $k = u$, the latter is precisely the constraint proven in Ref.~\cite{Bra14}.
The above derivation shows that this relation between the response coefficients arises directly from the non-zero lower bound on the entropy production in terms of the heat input rate Eq.~\eqref{input-bound-magnetic}.
The ensuing linear response power-efficiency tradeoff relation is
\begin{align}
\overline{\dot{W}} \leq \chi_t T_\text{c} \eta (\eta_\text{C}-\eta) \label{tradeoff-linear} ,
\end{align}
with the coefficient $\chi_t$ depending only on the temperature protocol $\phi(t)$.
This tradeoff relation has been derived from the constraint \eqref{onsager-bound-3} on the Onsager coefficients in Ref.~\cite{Bra14}.
Assuming that $L_{t u} \neq 0$, such that any variation in the force leads to heat flow, we define $\alpha_k = L_{k t}/L_{t k}$ and $\beta_k = L_{tk}^2/(L_{tt} L_{kk})$.
Thus $\alpha_k$ measures the asymmetry between the off-diagonal coefficients, with $\alpha_k=1$ corresponding to a symmetric Onsager matrix, whereas $\beta_k$ measures the overall magnitude of the off-diagonal coefficients relative to the diagonal ones, with $\beta_k = 0$ corresponding to a diagonal Onsager matrix.
The inequality \eqref{onsager-bound-3} then translates into a range of allowed values for $\alpha_k$ and $\beta_k$,
\begin{align}
\beta_k \leq \frac{1-\frac{1}{\gamma_t}}{\frac{1}{4}(\alpha_k-1)^2 + \alpha_k \Big(1-\frac{1}{\gamma_t} \Big)} \label{offdiag-constraint} ,
\end{align}
where we defined $\gamma_t \equiv \chi_t/L_{tt} \geq 1$. 
This constraint is tighter than the one obtained from the Second Law,
\begin{align}
\beta_k \leq \frac{4}{(1+\alpha_k)^2}
\end{align}
and in particular we always have $\beta_k \leq \gamma_t$, which limits the overall size of the off-diagonal Onsager coefficients for heat flows.
The two constraints on $\alpha_k$ and $\beta_k$ are compared in Fig.~\ref{fig:onsager}.

\section{Non-thermal dynamics} \label{sec-nonthermal}
In all the examples discussed in the previous Sections, the system was in contact with a thermal reservoir, allowing direct identification of the medium part of the entropy production in terms of a heat current.
In the following, we discuss what happens for non-thermal systems, where this connection is lost.
In order to have some notion of heat, we consider a particle with position $\bm{x}$ and velocity $\bm{v}$, with Hamiltonian $H(\bm{x},\bm{v},t) = m \bm{v}^2/2 + U(\bm{x},t)$.
As in the case of coupling to a heat bath, we assume that in addition to the force due to the potential $-\bm{\nabla}U(\bm{x},t)$, the particle is subject to a friction force and a stochastic force.
However, both the friction force and the stochastic force may now depend on both the position and the velocity of the particle, specifically
\begin{align}
\dot{\bm{v}} &= -\frac{1}{m} \bm{\nabla} U(\bm{x},t) - \gamma(\bm{x},\bm{v},t) \bm{v} + \sqrt{2 D(\bm{x},\bm{v},t)} \odot \bm{\xi} \label{langevin-non-thermal} \\ 
&= -\frac{1}{m} \bm{\nabla} U(\bm{x},t) - \gamma(\bm{x},\bm{v},t) \bm{v} + \bm{\nabla}_{\bm{v}} D(\bm{x},\bm{v},t) \nn
&\hspace{3 cm} + \sqrt{2 D(\bm{x},\bm{v})} \cdot \bm{\xi} \n .
\end{align}
In a change from Eq.~\eqref{langevin}, we now use the anti-It{\=o} stochastic integral, however, this can be brought into It{\=o} form by the straightforward transformation given in the second line.
For the special case of constant friction coefficient $\gamma(\bm{x},\bm{v},t) \equiv \gamma$ and the Stokes-Einstein relation $D(\bm{x},\bm{v},t) \equiv \gamma T(t)/m$, this reduces to the previously studied case of coupling to a thermal bath at temperature $T$ with Stokes-friction.
By contrast, a velocity-dependent friction and diffusion coefficient may occur as a result of the effective description of a non-thermal system.
Important examples include dry friction between surfaces of solids \cite{Gen05,Tou10,Bau11}, diffusion of cold atoms in dissipative optical lattices \cite{Dal89,Coh90,Dec15}, relativistic Brownian motion \cite{Dun09} and models of active particles \cite{Sch98}.
In all these cases, the friction and diffusion coefficient are even functions of the velocity $\gamma(\bm{x},-\bm{v},t) = \gamma(\bm{x},\bm{v},t)$ and $D(\bm{x},-\bm{v},t) = D(\bm{x},\bm{v},t)$, and we will assume this in the following.
Under this assumption, the reversible and irreversible probability currents are given by
\begin{align}
J_{x_i}^\text{rev}(\bm{x},\bm{v},t) &= v_i P(\bm{x},\bm{v},t), \qquad J_{x_i}^\text{irr}(\bm{x},\bm{v},t) = 0, \nn
J_{v_i}^\text{rev}(\bm{x},\bm{v},t) &= -\frac{1}{m} \big[\partial_{x_i} U(\bm{x},t)\big] P(\bm{x},\bm{v},t), \nn
J_{v_i}^\text{irr}(\bm{x},\bm{v},t) &= -\Big( \gamma(\bm{x},\bm{v},t) v_i + D(\bm{x},\bm{v},t) \partial_{v_i}  \Big) P(\bm{x},\bm{v},t) .
\end{align}
Since we chose the anti-It{\=o} interpretation in Eq.~\eqref{langevin-non-thermal}, the diffusion coefficient now appears in front of the velocity-derivative.
The change in the average energy of the particle is given by
\begin{align}
\dif{t}\av{H}_t &= -\dot{W}(t) + \dot{Q}(t) \quad \text{with} \\
\dot{W}(t) &= - \av{\partial_t U}_t \nn
\dot{Q}(t) &= m \Ind{x} \Ind{v} \ \bm{v} \cdot \bm{J}_{\bm{v}}^\text{irr}(\bm{x},\bm{v},t) \n.
\end{align}
Because the Hamiltonian is quadratic in the velocity, the system still permits decomposing the change in energy into a work and a heat contribution, and the latter is determined by the irreversible velocity probability current, just as for a thermal bath.
Just as before, the total entropy production rate Eq.~\eqref{entropy-current} is given by the square of the irreversible probability current
\begin{align}
\sigma^\text{tot}(t) = \Ind{x} \Ind{v} \ \frac{\big(\bm{J}_{\bm{v}}^\text{irr}(\bm{x},\bm{v},t)\big)^2}{D(\bm{x},\bm{v},t) P(\bm{x},\bm{v},t)} .
\end{align}
The medium part of the entropy production rate Eq.~\eqref{entropy-med} now reads
\begin{align}
\sigma^\text{med}(t) = -\Ind{x} \Ind{v} \ \frac{\gamma(\bm{x},\bm{v},t)}{D(\bm{x},\bm{v},t)} \bm{v} \cdot \bm{J}_{\bm{v}}^\text{irr}(\bm{x},\bm{v},t).
\end{align}
This is proportional to the heat only if the ratio $\gamma(\bm{x},\bm{v},t)/D(\bm{x},\bm{v},t)$ is independent of $\bm{x}$ and $\bm{v}$.
In this case, we define $T(t) \equiv m D(\bm{x},\bm{v},t)/\gamma(\bm{x},\bm{v},t)$, which can be understood as a generalized Stokes-Einstein relation \cite{Dub09}.
Then, we recover the relation $\sigma^\text{med}(t) = -\dot{Q}(t)/T(t)$, which thus holds for a thermal bath, irrespective of whether the friction force is linear in the velocity or not.
Even if such a relation does not hold, the heat flow still has the form of an irreversible generalized current \eqref{currents-rev-irr} and is thus bounded by
\begin{align}
\big(\dot{Q}(t)\big)^2 \leq m^2\av{D \bm{v}^2}_t \sigma^\text{tot}(t) .
\end{align}
In the particular case of a non-equilibrium steady state, we then have
\begin{align}
\big( \dot{Q} \big)^2 \leq m^2 \av{D \bm{v}^2}_\text{s} \sigma^\text{med} .
\end{align}
Thus, even though there is no relation between the heat flow and the entropy production rate in the form of an equality for non-thermal systems, the inequality relating the entropy production rate to the magnitude of the heat flow is preserved for the Hamiltonian plus non-equilibrium-bath dynamics discussed above.
Since the heat flow cannot be straightforwardly expressed in terms of the entropy production rate, the efficiency cannot be readily defined.
However, by introducing the effective temperature
\begin{align}
T^\text{eff}(t) \equiv \av{\theta}_t \quad \text{with} \quad \theta(\bm{x},\bm{v},t) = m\frac{D(\bm{x},\bm{v},t)}{\gamma(\bm{x},\bm{v},t)},
\end{align}
we can write the medium entropy production rate as
\begin{align}
\sigma^\text{med}(t) = -\frac{1}{T^\text{eff}(t)} \big(\dot{Q}(t) - \dot{\mathcal{Q}}(t)\big) .
\end{align}
The quantity $\theta(\bm{x},\bm{v},t)$ is a state-dependent \enquote{temperature}, whose average is the effective temperature $T^\text{eff}(t)$.
The above relation defines the non-thermal heat flow
\begin{align}
\dot{\mathcal{Q}}(t) = m\Ind{x} \Ind{v} \ \bigg( 1 - \frac{T^\text{eff}(t)}{\theta(\bm{x},\bm{v},t)} \bigg) \bm{v} \cdot \bm{J}^\text{irr}_{\bm{v}}(\bm{x},\bm{v},t) \label{heat-nonthermal}.
\end{align}
This non-thermal heat flow vanishes whenever the generalized Stokes-Einstein relation holds and thus $T^\text{eff}(t)$ is the actual temperature of the thermal heat bath.

As an example, consider a spatially periodic potential in one dimension and, as in Sec.~\ref{sec-ratchets-under}, the work output rate $\dot{W} = -F_0 \av{v}_t$ with the constant load force $F_0$.
Such a situation may be encountered in a ratchet driven by a non-linear friction force \cite{Bau12,Sar13,Sar13B}.
In the steady state, we then have $\dot{Q} = \dot{W}$ and we interpret $\dot{\mathcal{Q}}$ as the part of heat flow due to the non-thermal nature of the bath.
Since $\sigma^\text{med} \geq 0$, positive output work requires $\dot{\mathcal{Q}} \geq 0$.
We then define the efficiency as
\begin{align}
\eta = \frac{\dot{W}}{\dot{\mathcal{Q}}} = 1 - \frac{T^\text{eff} \sigma^\text{med}}{\dot{\mathcal{Q}}}.
\end{align}
Since the non-thermal heat flow is an irreversible generalized current, it is bounded by the entropy production rate
\begin{align}
\big(\dot{\mathcal{Q}}\big)^2 &\leq \chi (T^\text{eff})^2 \sigma^\text{med} \quad \text{with} \\
&\chi = \frac{m}{T^\text{eff}} \av{\gamma \theta \bigg(1-\frac{T^\text{eff}}{\theta} \bigg)^2 v^2}_\text{s} \n .
\end{align}
This then leads to the tradeoff relation
\begin{align}
\dot{W} \leq \chi T^\text{eff} \eta \big(1-\eta\big) .
\end{align}
Just as for a thermal system, the extracted power vanishes as the efficiency approaches its maximum value, which is in this case unity, since there is a single non-thermal bath.
However, it should be noted that the quantity $\dot{\mathcal{Q}}$ is only formally a heat flow, and may not be a measurable quantity or represent the true energetic cost of maintaining the bath in its non-equilibrium state.
Whether the above definition of efficiency is suitable thus depends on the specific system, and it generally does not correspond to a thermodynamic efficiency.
To clarify this point, let us further specify the dynamics.
We consider an overdamped particle in one dimension, subject to a spatially periodic potential $U(x+L) = U(x)$ and temperature profile $\theta(x+L) = \theta(x)$.
We further apply a constant load force $F_0$.
This situation is a ratchet model first discussed by B{\"u}ttiker \cite{Bue87} and Landauer \cite{Lan88}.
Similar to Sec.~\ref{sec-feynman-over}, we have the work rate $\dot{W} = - F_0 v$, where $v$ is the steady state drift velocity, and the heat flow
\begin{align}
\dot{Q} = \int_0^L \text{d}x \ \big[\partial_x U(x) - F_0 \big] J_\text{s} = \dot{W} \label{heat-BL} ,
\end{align}
since in one dimension the steady state current is independent of $x$.
The medium entropy production rate is
\begin{align}
\sigma^\text{med} &= -\int_0^L \text{d}x \ \frac{1}{\theta(x)} \big[\partial_x U(x) - F_0 + \partial_x \theta(x) \big] J_\text{s} \nn
&=-\int_0^L \text{d}x \ \frac{1}{\theta(x)} \big[\partial_x U(x) - F_0 \big] J_\text{s} \label{entropy-BL},
\end{align}
since the last term in the first line vanishes, being a total derivative of $\ln \theta(x)$.
Comparing Eqs.~\eqref{heat-BL} and \eqref{entropy-BL}, and defining $T_\text{eff} = \av{\theta}_\text{s}$, the non-thermal heat flow is given by
\begin{align}
\dot{\mathcal{Q}} = \int_0^L \text{d}x \ \bigg(1 - \frac{T^\text{eff}}{\theta(x)} \bigg) \big[\partial_x U(x) - F_0 \big] J_\text{s}.
\end{align}
Similar to Sec.~\ref{sec-periodic-under}, we parameterize the inverse temperature as
\begin{align}
\frac{1}{\theta(x)} = \frac{1}{T_\text{c}} + \bigg(\frac{1}{T_\text{h}} - \frac{1}{T_\text{c}} \bigg) \phi(x),
\end{align}
where the function $0 \leq \phi(x) \leq 1$ is now periodic in space $\phi(x+L) = \phi(x)$.
Using this, the entropy production rate can be split into two contributions in two different ways
\begin{align}
\sigma^\text{med} &= \frac{1}{T^\text{eff}} \big(- \dot{W} + \dot{\mathcal{Q}} \big) \\
&= -\frac{1}{T_\text{c}} \dot{W} + \bigg(\frac{1}{T_\text{c}} - \frac{1}{T_\text{h}} \bigg) \dot{Q}^\text{in} \n .
\end{align}
where the heat flows $\dot{Q}^\text{in}$ and $\dot{\mathcal{Q}}$ are defined as
\begin{subequations}
\begin{align}
\dot{Q}^\text{in} &= \int_0^L \text{d}x \ \phi(x) \big[\partial_x U(x) - F_0 \big] J_\text{s} \\
\dot{\mathcal{Q}} &= \bigg(1-\frac{T^\text{eff}}{T_\text{c}} \bigg) \dot{W} + \frac{T^\text{eff}}{T_\text{c}} \eta_\text{C} \dot{Q}^\text{in} .
\end{align}
\end{subequations}
While the work rate $\dot{W}$ is uniquely defined as the work per time performed against the external load, which of the two quantities $\dot{Q}^\text{in}$ and $\dot{\mathcal{Q}}$ is interpreted as the cost associated with the performed work depends on the physical setting.
Taking $\dot{Q}^\text{in}$ as the heat cost is the thermodynamic viewpoint that one has a cold reservoir at temperature $T_\text{c}$ and heat is absorbed from the hot reservoir, whose temperature $T(x) = \theta(x) > T_\text{c}$ in this case depends on the position of the particle.
In this situation, the efficiency is indeed bounded by the Carnot efficiency.
On the other hand, taking $\dot{\mathcal{Q}}$ as the heat cost corresponds to interpreting $\theta(x)$ as a single, out-of-equilibrium heat bath at effective temperature $T^\text{eff}$.
Note that since $T^\text{eff} > T_\text{c}$, the non-thermal heat flow is always smaller than the thermodynamic heat absorbed from the hot bath $\dot{\mathcal{Q}} < \dot{Q}^\text{in}$.
In this situation, the efficiency is not bounded by the Carnot efficiency and can in principle reach unity.

\section{Discussion}

The results for specific systems derived in this work are based on the general bound \eqref{current-inequality}, which we interpret as a quantitative statement of the Second Law of thermodynamics: It gives a positive and finite lower bound on the total rate of entropy production in terms of the square of any irreversible current in the system. 
Given the universality of the Second Law of thermodynamics, it is encouraging that its quantitative refinement can have similarly far-reaching consequences, imposing universal limits on the performance of engines in contact with a heat bath.
The preceding results also support the statistical definition of entropy production \eqref{path-entropy} in terms of the path probability---even though this quantity coincides with the thermodynamic definition of entropy only for thermal systems, it nevertheless continues to serve as a measure of irreversibility and provides a tangible upper bound on observable currents even for non-thermal systems.

For Langevin dynamics, the bound \eqref{current-inequality} follows from a simple application of the Cauchy-Schwarz inequality and is a consequence of the mathematical structure of the system, including the expression for the entropy production rate Eq.~\eqref{entropy-current}.
Since a similar bound, albeit for a more restrictive class of observables, was derived for a Markov jump process in Ref.~\cite{Shi16}, the more general bound may also hold for other observables in the latter case.

Finally, let us remark on the relation of our results to a family of recently derived bounds on stochastic currents in terms of their variance, referred to as thermodynamic uncertatinty relation \cite{Bar15,Pie17,Dec17}.
The latter implies that the proportionality constant between the current and the entropy production rate in Eq.~\eqref{current-inequality} should be related to the variance of the current.
Indeed, the bound \eqref{current-inequality} can be obtained as a short-time-limit of an inequality involving the variance of the current \cite{Dec17B}, however, the thermodynamic uncertainty relation itself only holds for the steady state of a dynamics that is even under time reversal.
Under these conditions, the bound Eq.~\eqref{current-inequality} is less tight than the uncertainty relation, however, it has the advantage that remains valid in the presence of time-dependent driving and odd variables under time-reversal.

\begin{acknowledgments}
\textbf{Acknowledgments.} The present study was supported by KAKENHI (Nos. 25103002, 17H01148 and 15F15324).
A.~D.~was employed as an International Research Fellow of the Japan Society for the Promotion of Science.
The authors wish to thank N. Shiraishi for stimulating discussions.
\end{acknowledgments}

\bibliography{bib}

\end{document}